\documentclass[amsmath,amssymb,floatfix,prd,12pt]{revtex4}
\usepackage{natbib}
\usepackage{graphicx}
\usepackage{bm}
\usepackage{amsmath}
\usepackage{hyperref}

\begin{document}

\title{$N$-Body Simulations of Alternative Gravity Models}
\author{Hans F. Stabenau}
\email{hstabena@physics.upenn.edu}
\author{Bhuvnesh Jain}
\email{bjain@physics.upenn.edu}
\affiliation{University of Pennsylvania\\
  209 S 33rd St, Philadelphia, PA 19104}

\pacs{04.50.+h}

\newcommand{\vecr}{{\bf r}}
\newcommand{\vecx}{{\bf x}}
\newcommand{\veck}{{\bf k}}
\newcommand{\vectheta}{{\bm \theta}}
\newcommand{\vecl}{{\bm \ell}}
\newcommand{\simgt}{\lower.5ex\hbox{$\; \buildrel > \over \sim \;$}}
\newcommand{\simlt}{\lower.5ex\hbox{$\; \buildrel < \over \sim \;$}}

\begin{abstract}
  Theories in which gravity is weaker on cosmological scales have been
  proposed to explain the observed acceleration of the
  universe.  The nonlinear regime in such theories is not well 
  studied, though it is likely that observational tests of structure
  formation will lie in this regime. A class of alternative gravity 
  theories may be approximated by modifying Poisson's equation. 
  We have run N-body simulations of a set of such
  models to study the nonlinear clustering of  matter on 1--100~Mpc
  scales. We find that nonlinear gravity enhances the
  deviations of the power spectrum of these models from standard gravity. This
  occurs due to mode-coupling, so that models with an excess or
  deficit of large-scale power (at $k < 0.2$~Mpc$^{-1}$) lead to
  deviations in the power spectrum at smaller scales as well (up to $k
  \sim 1$~Mpc$^{-1}$), even though the linear spectra match very
  closely on the smaller scales. This  makes it
  easier to distinguish such models from general relativity using
  the three-dimensional power spectrum probed by galaxy surveys and
  the weak lensing power spectrum.
  If the potential for light deflection is modified in the same
  way as the potential that
  affects the dark matter, then weak lensing constrains deviations
  from gravity even more strongly.
  
  Our simulations show that even with a modified potential,
  gravitational evolution is approximately universal, and our
  conclusions extend to models with modifications on scales of
  1--20~Mpc.  Based on this, the Peacock-Dodds approach can be adapted
  to get an analytical fit for the nonlinear power spectra of
  alternative gravity models, though the recent Smith et al formula is
  less successful. We also use a way of measuring projected power
  spectra from simulations that lowers the sample variance, so that
  fewer realizations are needed to reach a desired level of accuracy.

\end{abstract}

\maketitle

\section{Introduction}
\label{sec:intro}

Observations of Type Ia supernovae, along with observations of the
cosmic microwave background and large-scale structure, have
established that the expansion of the universe is accelerating
\citep{Knop:2003iy,Spergel:2003cb,Riess:2004nr}.  
Einstein's theory of gravity, and a
cosmological model that includes dark matter, baryons and radiation,
cannot explain this cosmic acceleration.  The explanation may involve
the existence of an exotic form of energy density, or the breakdown of
general relativity (GR) on large scales.  These explanations for cosmic
acceleration are known as ``dark energy'' and ``alternative gravity''
approaches, respectively.

The rate at which the universe expands is predicted by the Friedmann
equation  $H^2 = 8 \pi G \rho/3$ (for a spatially flat universe), 
which is derived from the Einstein
equations and the metric.  In order to reproduce the cosmic
acceleration, we have to modify the Einstein equation:
\begin{equation}
  \label{eq:Einstein}
  R_{\mu\nu}-\frac{1}{2}R g_{\mu\nu} = 8 \pi G T_{\mu\nu}.
\end{equation}
The left hand side describes the curvature of spacetime due to gravity
while the right hand side describes its sources.  Given an observed
expansion history that one wishes to describe, an alternative gravity
(AG) theory will attempt to explain it via a modification of the
left-hand side while a dark energy (DE) theory introduces a new term
on the right-hand side that gives the desired acceleration.  While one could
imagine two different kinds of modifications that gave the same
expansion history, they will have different effects on the growth of
structure by gravitational clustering: a smooth dark energy (DE)
affects the growth of structure only by changing the expansion rate,
while AG affects it by direct modification of the gravitational
interaction.  The linear regime growth factor $G(a)$ is scale
independent in DE models; an AG modification will produce a different
solution, which in general is scale-dependent: $G(k,a)$ as described
in Sec.~\ref{sec:linear_regime} (though see \citet{Jacobs:1993tm} for
weak scale dependence in standard gravity due to the effect of
inhomogeneities).

There are several potential ways of modifying GR:
adding nonlinear terms in the Ricci scalar $R$ to the gravitational
action, coupling $R$ to a scalar field as in Brans-Dicke gravity,
having gravity operate in a higher dimensional universe on large
scales as suggested by brane cosmology, introducing scalar and vector
degrees of freedom and so on. The tensor-vector-scalar theory of
\citet{Bekenstein:2004ne}, the ghost condensate theory of
\citet{Arkani-Hamed:2003uy}, and the five-dimensional theory of
\citet*[DGP gravity]{Dvali:2000qq} have drawn recent
attention. Our interest is in the class of theories such as DGP that
aim to reproduce cosmic acceleration by weakening gravity on
large-scales. The theory is then likely to have testable consequences
on scales probed by large-scale structure. 

\citet{Lue:2004rj} have derived the linear growth of
perturbations in DGP gravity. Further, \citet{Lue:2003ky} 
argued that generic gravity theories that obey Birkhoff's theorem and 
mimic cosmic acceleration lead to
the suppression of the growth of large-scale density perturbations
at the level of $\sim 5\%$, similar to DGP.  
Predictions for the exact linear growth
as a function of scale, or for scale-dependent growth in the
nonlinear regime, do not yet exist in such theories.  

Newton's Law has been directly measured from millimeter to solar
system scales \citep{Adelberger:2003zx, Hoyle:2004cw}.  To constrain
possible deviations from Eq.~(\ref{eq:Einstein}) on cosmological
scales requires geometric information (distance measurements to
objects of known redshift) and information on the growth of structure.
Since the observed cosmic acceleration occurs at low redshift,
observational constraints at $z<1$ are needed to learn about its
origin. Geometric information at low redshift has been measured most
cleanly by the measurement of the luminosity distances $d_L(z)$ to
type Ia supernovae. Weak lensing (WL) can measure low-redshift
distance information as well as the evolution of the growth factor,
especially with tomography \citep{Hu:1999ek}. Baryon acoustic
oscillations in the power spectrum of galaxies or other probes can
measure the angular diameter distance at low redshifts.  The CMB
(cosmic microwave background) measures geometric information (the
angular diameter distance to last scattering at $z=1089$), the shape
and amplitude of the primordial power spectrum, and matter/radiation
content \citep{Dodelson:2003}.  Thus the CMB anchors the cosmological
model at high redshift, and by comparing to it, Type Ia SN, weak
lensing, baryon oscillations and galaxy cluster measurements constrain
the effects of dark energy or AG.  Currently, combining the CMB, SN,
and large-scale structure information has led to a best-fit
``standard'' cosmological model \citep{Spergel:2003cb}. This model is
usually described in terms of standard gravity (GR) and dark energy;
current constraints on the equation of state of dark energy are
consistent with a cosmological constant, but a possible time evolving
dark energy is not well constrained.

Some studies have considered the constraints on alternative gravity from
existing and planned survey data \citep[e.g.][]{Liddle:1998ij,
  Uzan:2000mz, Peebles:2002iq,
  Song:2005gm,
  Song:2006sa, Knox:2005rg, Ishak:2005zs, Alam:2005pb,
  Sawicki:2005cc,Koyama:2006ef,
  Maartens:2006yt}.
\citet{Lue:2004rj} and some of these authors
have examined the linear regime growth factor in DGP gravity and
computed its consequences for low-redshift galaxy power spectra and 
weak lensing observables. 

In this paper we study the consequences of modifying
Poisson's equation for the growth of structure in the nonlinear
regime. The model we consider
for the modified Poisson equation may be regarded as an approximate
description of a complete AG theory over a range of scales (though not
all AG theories will be described by our approach to gravitational 
clustering). While our AG model
is not derivable from a covariant, consistent theory of gravity, it
has the merit that we can use N-body simulations to study the 
the full nonlinear evolution of structure. We are
interested in the amplitude and scale dependence of modified growth
over the range 1--100~Mpc, where cosmological observations can
probe gravity effectively. An AG theory that provides an 
expansion history similar to the $\Lambda$CDM accelerating 
universe is likely to alter Newton's Law on these scales. 
We note that a similar modification of the gravitational potential was
considered by \citet{White:2001kt}. They followed a somewhat 
different approach to
lensing (changing the deflection angle relation while retaining the
growth of structure as in standard gravity) and calculated the
consequences on smaller scales (0.01--10~Mpc), as their focus was on
modifications that produce flat rotation curves for galaxies without
the need for dark matter. 

In the linear regime, analytic
calculations can explore the effects of a modification of Poisson's 
equation on the growth of structure.  
Recent efforts in this direction were made by
\citet{Shirata:2005yr} and \citet{Sealfon:2004gz}. However observations have
significant information in the small scale nonlinear regime, so it is
necessary to develop predictions for this regime. Moreover, we know
from perturbation theory that quasilinear effects propagate power from
large to small scales, so altering gravity on large scales is likely
to affect smaller structure as well.  To obtain accurate nonlinear
predictions, we use N-body simulations to determine the effect of 
modifications to Poisson's equation in the non-linear regime
on structure formation. We quantify the extent to which such a 
modification would be constrained by galaxy and WL surveys.

In Sec.~\ref{sec:linear_regime} we describe the formalism that describes
the growth of perturbations due to gravity. Sec.~\ref{sec:nbsims}
contains details on our numerical simulations and predictions for
three-dimensional and lensing power spectra.  We describe our results
in Sec.~\ref{sec:sims} and conclude in Sec.~\ref{sec:discussion}.


\section{Linear Regime}
\label{sec:linear_regime}

In Eulerian coordinates the equations that govern the behavior of mass 
fluctuations are given by recasting the fluid equations in expanding 
coordinates, or simply by conservation of stress-energy
$\nabla_\mu T^{\mu\nu}=0$.  If we linearize these equations,
the resulting second order differential equation describes the 
growth of the fractional overdensity
$\delta({\bf r},t)$, or equivalently, its Fourier transform
$\tilde{\delta}({\bf k}, t)$:
  \begin{eqnarray}
    \label{eq:delta}
    \ddot{\delta} + 2H\dot{\delta} &=& \frac{\nabla^2
      \phi}{a^2},\\
    \label{eq:deltak}
    \ddot{\tilde{\delta}} + 2H\dot{\tilde{\delta}} &=& 
-\frac{k^2}{a^2}\tilde{\phi},
\end{eqnarray}
where $a(t)$ is the expansion scale factor and gives 
the Hubble parameter as $H(t) \equiv \dot{a}/a$.  The Fourier transformed Poisson equation in
comoving coordinates reads
\begin{equation}
  \label{eq:phik_cont}
  {\tilde\phi}(\veck,t) = 
-\frac{3}{2} \frac{H_0^2 \Omega_{\rm m0}}{a} 
\frac{{\tilde\delta}(\veck,t)}{|\veck|^2},
\end{equation}
where $k$ is the comoving wavenumber and $\phi$ is the gravitational
potential.  In this work our continuous Fourier transform conventions
are
\begin{eqnarray}
  \tilde{\delta}(\veck) &=& \int d\,^3 \vecr\, \delta(\vecr) e^{i \veck \cdot \vecr},\\
  \delta(\vecr) &=& \int \frac{d\,^3 \veck}{(2\pi)^3}\, \tilde{\delta}(\veck) e^{-i\veck\cdot\vecr}.
\end{eqnarray}
A DE modification will change Eq.~(\ref{eq:deltak}) via the time
derivatives and the Hubble parameter $H\equiv \dot{a}/a$ on the left
hand side; for such a model one separates Eq.~(\ref{eq:deltak}) by
letting $\tilde{\delta}(\veck,t) \equiv \tilde{\delta}(\veck) G(t)$
and then solving for the growth factor $G(t)$.  
An AG modification will change the potential on the right-hand side via
Eq.~(\ref{eq:phik_cont}).  We can see from this that if an AG
modification is made, i.e.\ if $k^2\phi({\bf k},t)\sim
\tilde{\delta}({\bf k},t)f(\veck,t)$ for some non-trivial $f(\veck,t)$,
Eq.~(\ref{eq:deltak}) will no longer be separable, and hence the
growth factor will become scale dependent, i.e.\ one must allow
$\tilde{\delta}(\veck,t) \equiv \tilde{\delta}(\veck)G(\veck,t)$.
Eq.~(\ref{eq:deltak}) then becomes
\begin{equation}
  \label{eq:Gk}
  \ddot{G}(\veck,t) + 2H\dot{G}(\veck,t) = 
  \frac{3}{2}\frac{H_0^2\Omega_{\rm m0}}{a^3} G(\veck,t) f(\veck,t).
\end{equation}
If $f(\veck,t)\rightarrow 1$ then $G(\veck,t)\rightarrow G(t)$ as in
standard gravity.  
We note that a purely time-dependent modification $f(t)$ (such as a
time-dependent Newton's constant), can be accommodated without a
scale-dependent growth factor. 

\subsection{Modified Poisson Equation}

\begin{figure}[t]
  \centering
  \includegraphics[width=0.6\linewidth]{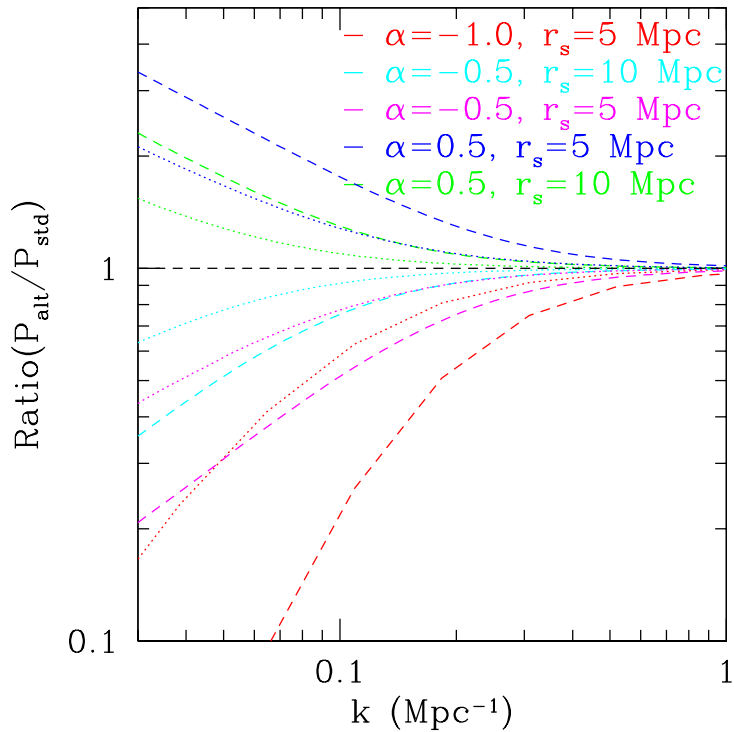}
  \caption{\small Ratios of linear theory alternative gravity (AG) 
    power spectra to
    standard gravity at $z=0$, for different parameterizations
    of Eq.~(\ref{eq:phikalt}).  The dotted lines are the matter power
    spectrum ratios $P_{\delta,{\rm alt}}/P_{\delta,{\rm std}} =
    (G_{\rm alt}(k,t)/G_{\rm std}(t))^2$.  The dashed lines are the
    ratios of the velocity power spectra $P_{v,{\rm alt}}/P_{v,{\rm
        std}} = (\dot{G}_{\rm alt}(k,t)/\dot{G}_{\rm std}(t))^2$.  The
    expansion history is fixed to be the same as for a $\Lambda$CDM
    universe.}
  \label{fig:ggdot}
\end{figure}

A theory of gravity that makes gravity weaker or stronger over a range
of length scales can be approximated by modifying Poisson's equation. 
\citet{Sealfon:2004gz} investigated the effect of a Yukawa-type
(adding an exponential term)  and
power-law modifications to the Poisson equation, solving for the
scale-dependent growth factor $G(\veck,t)$ under the assumption that
the modification was a small perturbation to the Newtonian potential.
\citet{Shirata:2005yr} followed by obtaining $G(\veck,t)$ for a Yukawa
modification of arbitrary strength.  They adopted the Peacock-Dodds
(PD) prescription \citep{Peacock:1996ci} --- an approach we check with
our simulations --- to obtain the non-linear matter power spectrum
from the linear solution. With a prescription for galaxy biasing, this
enabled them to predict galaxy power spectra, which they used with
their AG model to constrain model SDSS galaxy power spectrum
measurements \citep{Tegmark:2003uf}.

Both \citet{Shirata:2005yr} and \citet{Sealfon:2004gz} consider the real-space potential
\begin{equation}
  \label{eq:phialt}
  \phi_{\rm alt}(\vecr) = 
-G\int d\,^3 \vecr' \frac{\rho(\vecr')}{|\vecr-\vecr'|}
  \left[1+\alpha\left(1-e^{-\frac{|\vecr-\vecr'|}{r_{\rm s}}}\right)\right].
\end{equation}
In Fourier space, this becomes
\begin{equation}
  \label{eq:phikalt}
  \widetilde{\nabla^2\phi}_{\rm alt}(\veck) =
  \frac{3}{2}\frac{H_0^2 \Omega_{\rm m0}}{a}\delta(\veck) \left[1 + \alpha\frac{1}{1
      + (|\veck| r_{\rm s}/a)^2}\right].
\end{equation}
Note that this will result in a scale dependent growth factor when
plugged into Eq.~(\ref{eq:delta}) above.
In Fig.~\ref{fig:ggdot}, we can see the effect of this modification on
the linear theory power spectrum; there we have plotted the ratio of
the AG matter and velocity linear power spectra to the corresponding
standard linear spectrum at redshift $z=0$ for a few different
parameterizations of Eq.~(\ref{eq:phikalt}).  Throughout this work, we
fix the background expansion to be the same as $\Lambda$CDM, i.e.\ we
take
\begin{equation}
  \label{eq:H}
  H^2(a)=H_0^2 (\Omega_{\rm m0}a^{-3} + \Omega_{\rm \Lambda 0}).
\end{equation}
In this study we do not allow the acceleration to vary with $\alpha$
or $r_s$; we let it be fixed solely by $\Lambda$, because an AG theory
would need to predict an accelerating expansion not too different from
that given by $\Lambda$CDM in order to fit the supernova data.  For
Fig.~\ref{fig:ggdot} and our simulations we take $\Omega_{\rm
  m0}=0.3$, and $\Omega_{\rm \Lambda 0} = 0.7$.

The matter power spectrum
$P_\delta(k) \propto G^2(k,t)$, while the velocity power spectrum
$P_v(k) \propto \dot{G}^2(k,t)$, so the ratios of the growth factors 
give us the ratios of the linear spectra starting from the same
initial spectrum. We solve for the growth factor using
Eq.~(\ref{eq:Gk}). The velocity power spectra show a
more pronounced difference than the matter spectra (a factor of 2-5
larger deviation at $k\approx 0.05$ Mpc$^{-1}$). It may be worth
exploring the measurement of large-scale peculiar velocities 
via the kinetic SZ effect or distance measurements on galaxies 
to test gravity. 

On large and small scales, we get the limiting behavior
\begin{eqnarray}
r \gg r_{\rm s}, \ \ \phi_{\rm alt} &\rightarrow&
(1+\alpha)\phi_{\rm newton}, \\
r \ll r_{\rm s}, \ \ \phi_{\rm alt} &\rightarrow& \phi_{\rm newton}. 
\end{eqnarray}
So for positive $\alpha$, the alternative gravity potential is
stronger than Newtonian gravity on large scales and unchanged on small
scales.  \citeauthor{Shirata:2005yr} take $r_{\rm s}$ to be a fixed
physical length, that is, in comoving units it changes with redshift;
consequently, at early times when $a \ll 1$, $r_s$ becomes very large.
At our simulation starting point of $z=50$, $r_s$ is much larger than
the boxsize, and so the linear spectra are virtually identical.  Hence
both the alternative and the standard gravity simulations start from
the same initial conditions.  We examine the effect of the initial
conditions further in Sec.~\ref{sec:sims}.

%

We note that this modification cannot extend to arbitrarily large
scales or early times. We regard it as an approximate description of
gravity on length scales well below the horizon at low redshifts. 


%


\section{Numerical Simulations}
\label{sec:nbsims}

While Eq.~(\ref{eq:delta}) is useful for describing linear-regime density
fluctuations on large scales, if we want a more complete description
we have to turn to numerical simulations.  An N-body code simulates
the evolution of structure by evolving a large number of particles
interacting by gravity. These are evolved from early times ($z\gg 1$)
up to the present, with particle positions and velocities outputted at
regular intervals. 

An efficient N-Body solver must compute the forces on a large number
of particles simultaneously so that the equations of motion can be
integrated forward in time.  We use a basic particle-mesh (PM) solver
for this purpose, which interpolates the particles onto a grid and
then computes the potential via Fourier transform.  More advanced
techniques (e.g.\ P$^3$M and tree codes) are available, which provide larger
dynamic range in exchange for greater complexity and computation time.
We use PM simulations to simulate modified gravity in the quasilinear
to moderately nonlinear regime where observations
can test models without needing to consider astrophysical/baryonic
effects.  Due to the lower CPU costs of PM simulations, 
we were able to run a large number of realizations to
reduce sample variance on the power spectra, which would have been
prohibitive with the other methods.  Our code is based on the PM code
of \citet{Klypin:1997sk}, which was designed and tested for DM
simulations \citep{Klypin:1992sf,Klypin:1994iu}, and kindly made
available by A. Klypin.  We set up the
initial conditions by displacing particles from a regular grid using a
realization of the linear power spectrum. 

\subsection{Discrete Poisson Equation}
\label{sec:standard_gravity}

In the PM simulation, the equations of motion are discretized on
a grid, starting with the Poisson equation. Since we modify this
equation for the AG simulations, we give the explicit formulae here. 
We define the second derivative operator in 1-D as
\begin{displaymath}
  \nabla^2\phi_i \approx \phi_{i+1} + \phi_{i-1} -2\phi_i.
\end{displaymath}
We define the unnormalized discrete Fourier transform as
\begin{eqnarray*}
  \tilde{\phi}_k &=& \sum_{r=0}^{N-1} \phi_r e^{i2\pi rk/N},\\
  \phi_r &=& \sum_{k=0}^{N-1} \tilde{\phi}_k e^{-i2\pi rk/N}.
\end{eqnarray*}
Combining the previous two equations leads to the discrete Fourier
space expression
\begin{equation}
  \label{eq:del2_dft}
  \widetilde{\left[\nabla^2\phi\right]}_k = \tilde{\phi}_k \times
  2\left[\cos{\frac{2\pi k}{N}}-1\right]
\end{equation}
So the 3-D discrete Poisson equation for standard gravity in
Fourier space reads
\begin{equation}
  \label{eq:phik_dft}
  \tilde{\phi}_k = \frac{3}{2} \frac{H_0^2\Omega_{\rm m0}}{a} \frac{\tilde{\delta}_k}{G_k},
\end{equation}
where
\begin{equation}
  \label{eq:gk}
  G_k \equiv 2\left[\cos{\frac{2\pi k_x}{N}} + \cos{\frac{2\pi
        k_y}{N}} + \cos{\frac{2\pi k_z}{N}} - 3\right].
\end{equation}

\begin{figure*}[p]
  \centering
  \includegraphics[width=0.425\linewidth]{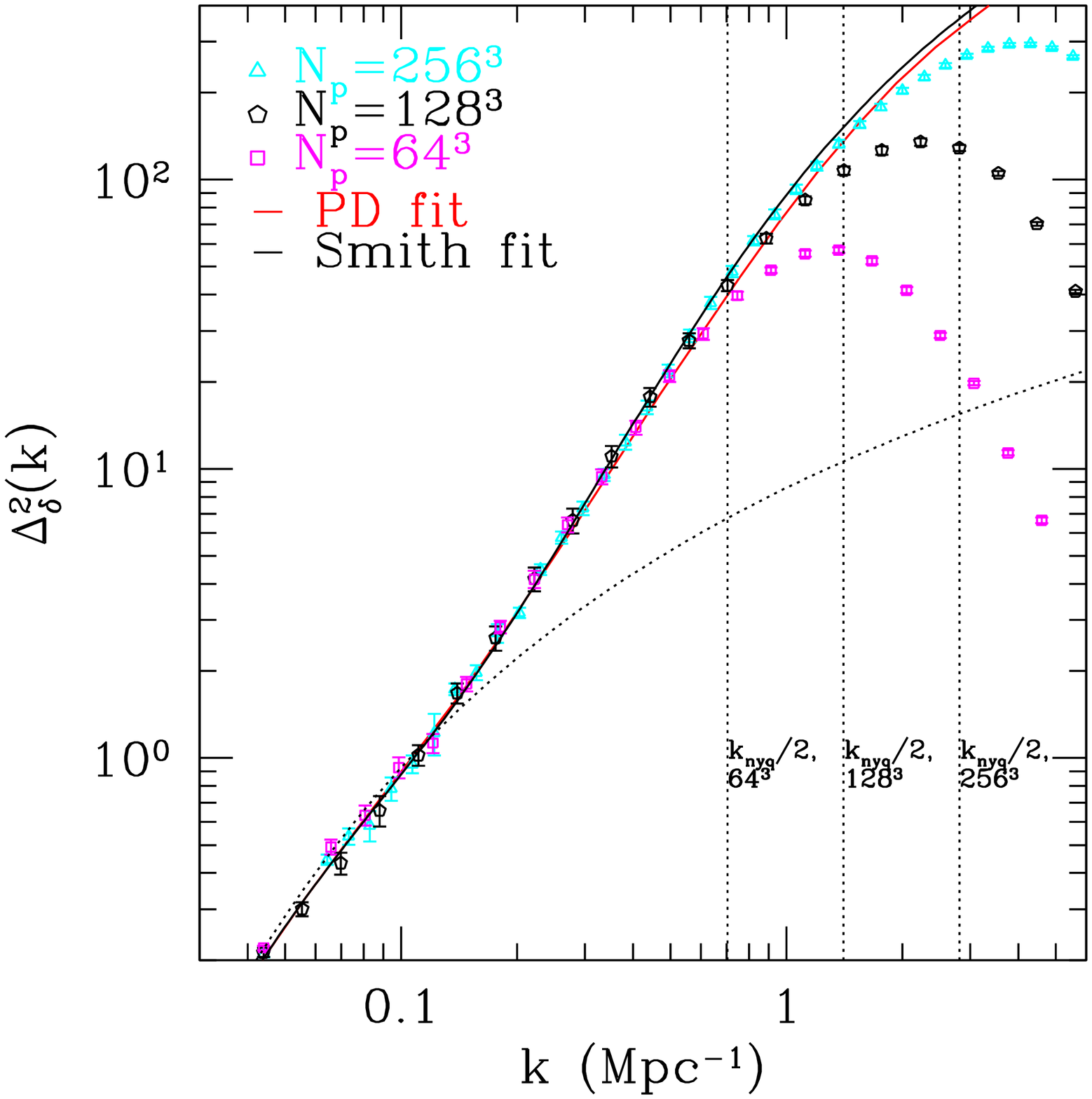}
  \hfil
  \includegraphics[width=0.425\linewidth]{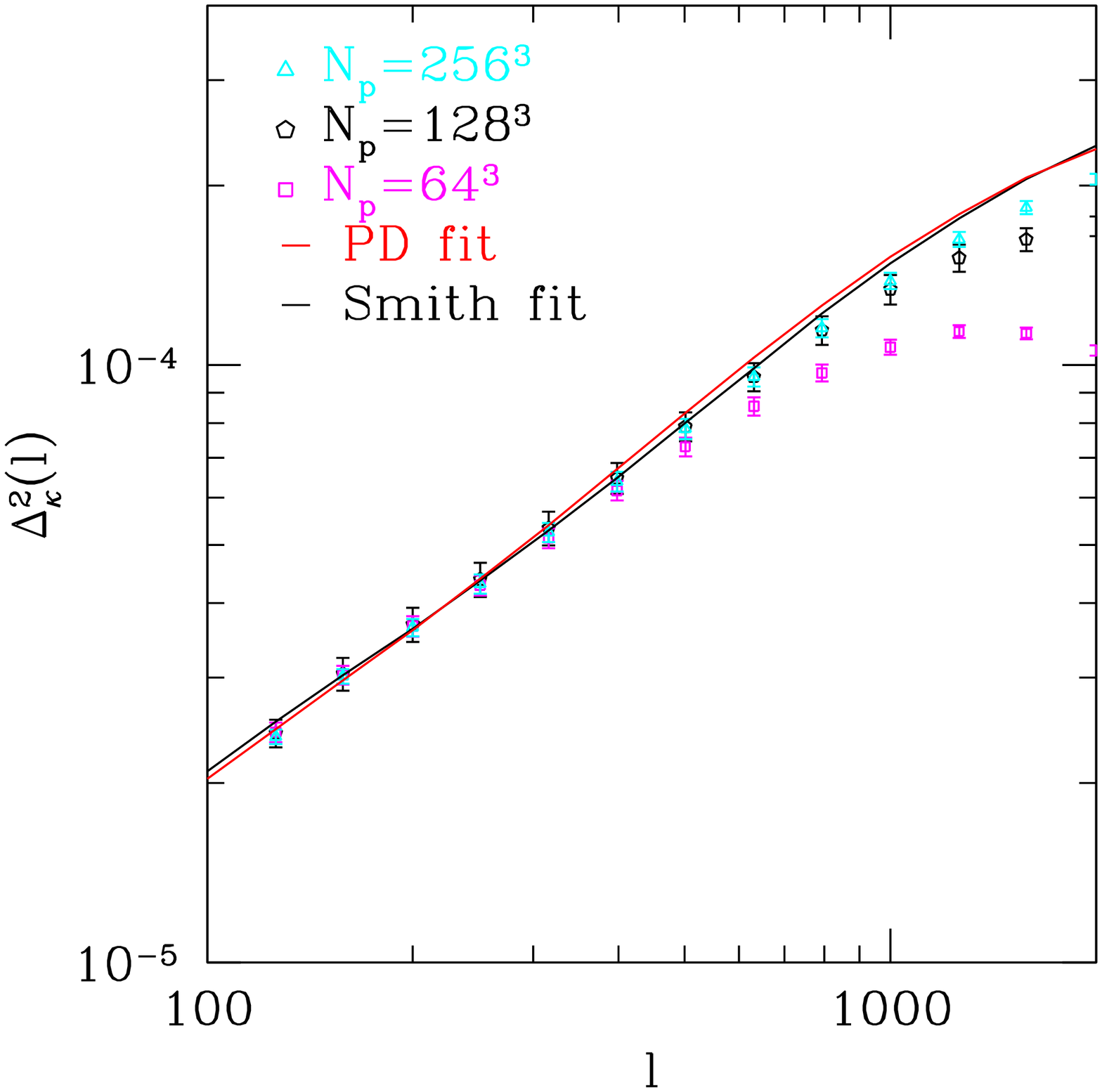}
  \caption{\small Dimensionless 3D power ($\Delta^2_\delta(k) =
    k^3P_\delta(k)/2\pi^2$, left panel) 
    and convergence power ($\Delta^2_\kappa(\ell) = 
   \ell^2P_\kappa(\ell)/2\pi$, right panel) 
   from N-body simulation with a standard
    (Newtonian) gravity model. Also shown are
    the Smith et al.\ (solid black line) and Peacock-Dodds
    (solid red line)
    fitting formulae.  The points with error bars are an average
    over 8 realizations. Three sets of simulations are shown with varying 
    resolution due to differences in the total number of particles; the total
    number of grid points in each set is $N_{\rm g}=8N_{\rm p}$.
  }
  \label{fig:standard}
\end{figure*}

\begin{figure*}[p]
  \centering
  \includegraphics[width=0.425\linewidth]{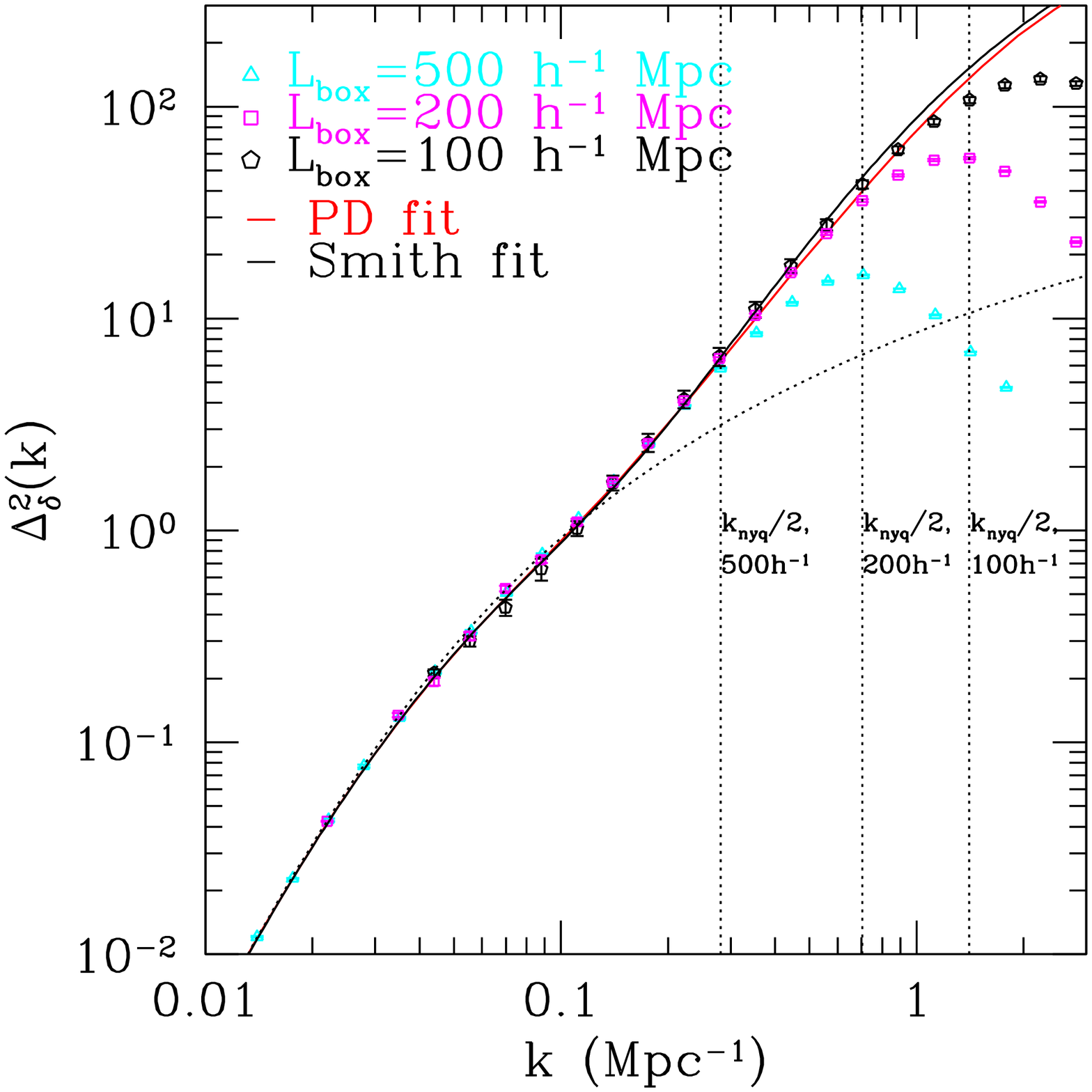}
  \hfil
  \includegraphics[width=0.425\linewidth]{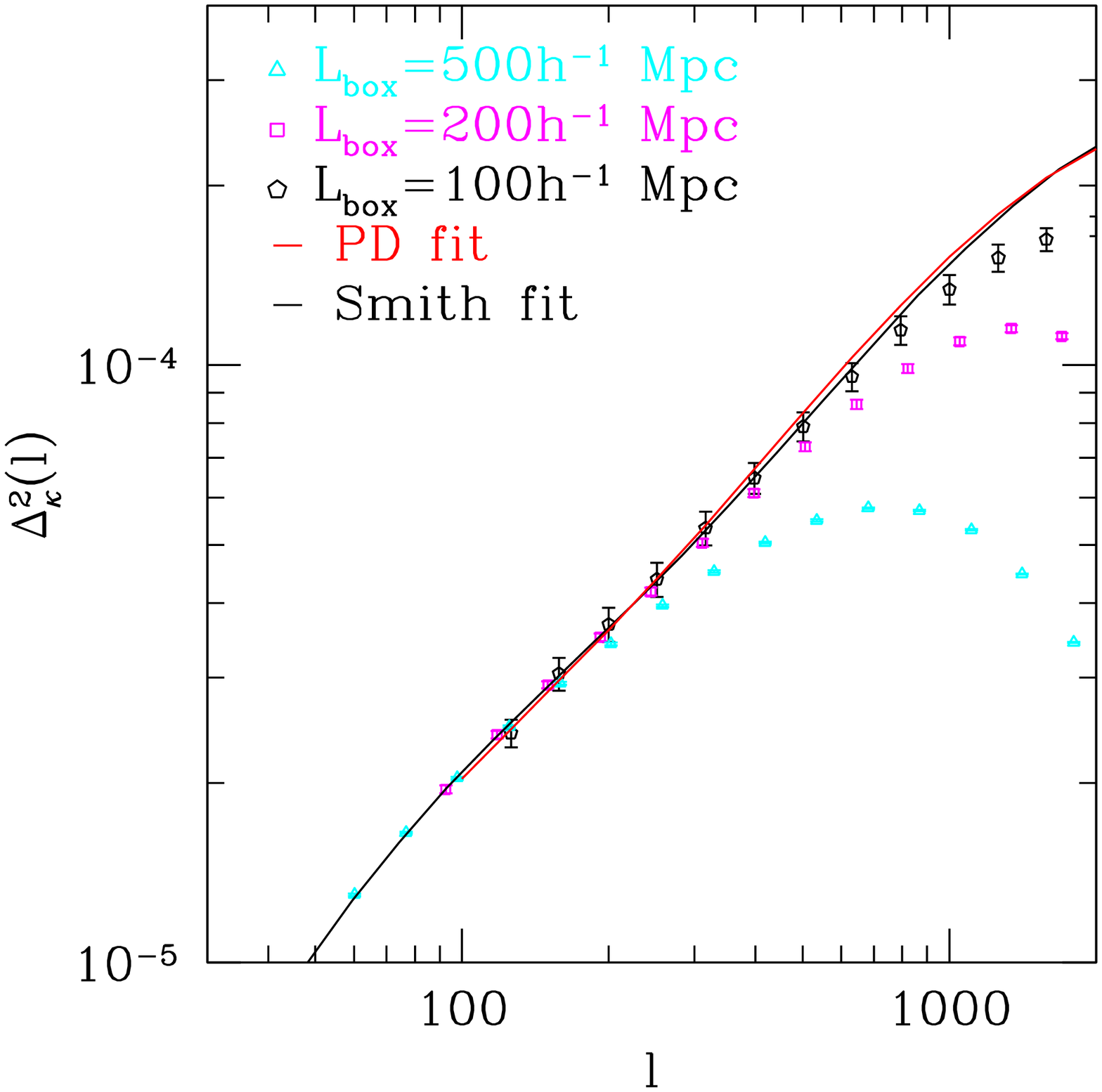}
  \caption{\small Dimensionless 3D power and convergence power spectra
    from simulation, shown together with the Smith and PD fitting formulae as
    in Fig.~\ref{fig:standard}.  Three different boxsizes are shown:
    $100h^{-1}$, $200h^{-1}$, and $500h^{-1}$ Mpc.}
  \label{fig:boxes}
\end{figure*}

\subsection{Particle-Mesh Simulation Parameters}
\label{sec:PM}

The parameters which determine the dynamic range of a PM simulation
are the boxsize $L_{\rm box}$, the number of particles $N_{\rm p}$,
and the Fourier grid size $N_{\rm g}$.  The grid
spacing should be smaller than the mean particle spacing to preserve
the small scale resolution; a common
choice which we adopt is to take $N_{\rm g} = 8N_{\rm p}$, i.e. twice
the number of particles per dimension.  
$L_{\rm box}$ must be large enough so that we have enough
power in the linear regime to get accurate power spectra, however
increasing $L_{\rm box}$ decreases one's ability to resolve
small-scale structure for fixed $N_{\rm p}$.  Our computational
resources fixed $N_{\rm g} = 256^3$ and $N_{\rm p} = 128^3$; we chose
$L_{\rm box}=100h^{-1}\simeq 140$~Mpc for our boxsize. 
The wavenumber  corresponding to $L_{\rm box}$ 
is $k_{\rm min} \approx 0.04$~Mpc$^{-1}$; the comoving
distance to the sources is $\chi(z = 1) \approx 3.3$~Gpc, so 
the angular wavenumber is $\ell_{\rm min}=k_{\rm min}\chi \approx 145$,
corresponding to a field of view $\sim 2.5^\circ$ on a side.

To check our results in the non-linear regime, and to insure that we
chose $L_{\rm box}$ large enough, we tested our prediction for the 3D power
for standard gravity against the \citet{Smith:2002dz} fitting formula.
We find in Fig.~\ref{fig:standard} that for our runs of $N_{\rm
  p}=128^3$ particles on a $N_{\rm g}=256^3$ grid, our results are
limited by resolution at physical scales of about 1.0~Mpc, and angular
scales of $\ell \approx 1000$.  The power spectrum $\Delta^2_\delta(k)$ in
Fig.~(\ref{fig:standard}) is identical to the linear theory prediction
on large scales; we have about a decade of power in the linear regime
at $z=0$.  
With our sources at $z_s = 1$, the weak lensing
weight function $W(\chi)$ in Eq.~(\ref{eq:lensing_W}) peaks at
$z\approx 0.4$ where the linear regime extends to smaller scales. So we can be
confident that for purposes of measuring the lensing power spectrum our
choice of $L_{\rm box} = 100h^{-1}$~Mpc is large enough.

Fig.~\ref{fig:standard} and Fig.~\ref{fig:boxes} show how the
resolution limit behaves as we change the number of particles or the
boxsize in the simulation.  We note that since we are not including
direct particle-particle effects (as in P$^3$M-type codes) we don't
need to be concerned with explicit force softening, and that the
shot-noise contribution is very small on the scales that we resolve.

The particle mass in our simulation $m_{\rm p}$ is given by
\begin{equation}
  \label{eq:mp}
  m_{\rm p} = \Omega_{\rm m0} \ \rho_{{\rm cr},0} \ 
\left(\frac{L_{\rm box}^3}{N_{\rm p}}\right)
\end{equation}
For our simulations with $N_{\rm p} = 128^3$ and $L_{\rm box} =
100h^{-1}$~Mpc, we get $m_{\rm p} = 1.1 \times 10^{10}$ $M_{\odot}$.

We ran all of our simulations in a ${\rm \Lambda CDM}$ background
cosmology.  They were started at redshift $z=50$. 
We used $\sigma_8 = 1.0$, $H_0 = 70$, $\Omega_\Lambda = 0.7$, and
$\Omega_m = 0.3$ as our cosmological parameters.  We ran 8
realizations for our runs with standard gravity and for each of the
alternative gravity models.

\subsection{Convergence Power Spectrum}
\label{sec:lenssim}

The output of N-Body simulations has been used to make model shear and
convergence maps of cosmological weak lensing
\citep{Jain:1999ir,White:1999xa}.
Unlike the angular power spectrum
of the CMB, which can be computed analytically, weak lensing involves
the deflection of light by large scale structure at low redshift,
which has already undergone significant nonlinear collapse, so
model WL maps and power spectra must be computed numerically.  

In the standard multiple lens plane algorithm, the convergence and
shear maps are computed by filling the light cone from the observer to
the source galaxies with matter from N-body 
simulation outputs. Starting at the source redshift, one can output three
orthogonal 2D-projections (slices) of the density field at length
intervals equal to the box size.  By picking a projection at each
redshift and randomly translating the boxes, we tile approximately
uncorrelated mass distributions along a light cone for each
realization. The box size at the source redshift determines the field
of view simulated.

\begin{figure}[t]
  \includegraphics[width=0.6\linewidth]{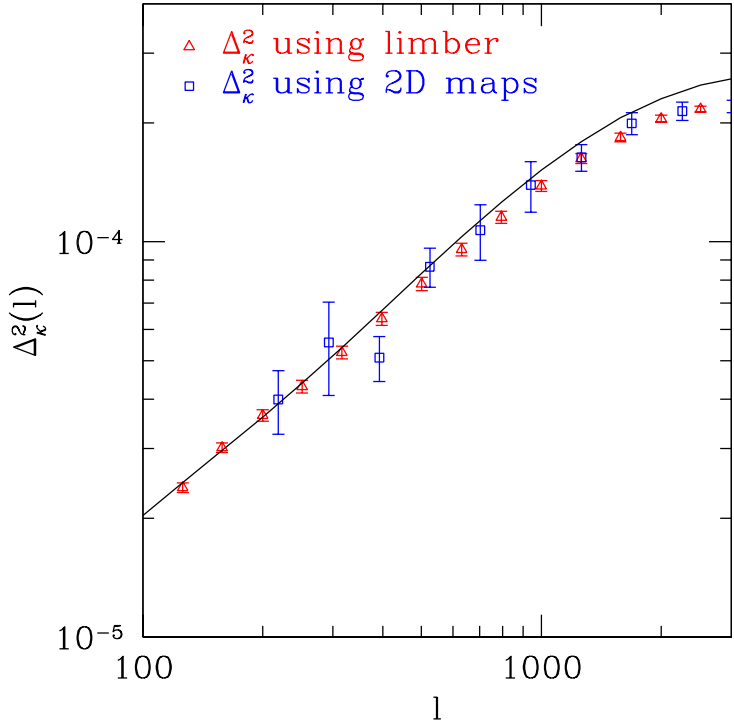}
  \caption{\small Comparison of the dimensionless convergence power 
    using our Limber approximation method (red
    triangles) and the standard multiple lens plane method (blue
    squares), each averaged over 8 realizations. Computing
    $P_\kappa$ in the standard method 
    (using Eq.~(\ref{eq:Pkappa_sum})) uses only the 2D
    modes of the density field at each redshift slice. 
    We included the full 3D modes by
    using the Limber equation for $P_\kappa$
    (Eq.~(\ref{eq:limber})). The scatter is
    significantly smaller, as shown by the error bars on the red
    symbols. The black curve
    is computed using the 
    \citet{Peacock:1996ci} 3D power spectrum. }
  \label{fig:2dmaps}
\end{figure}

Following \citet{Bartelmann:1999yn}, in the thin-lens approximation
for a flat universe, the convergence $\kappa(\theta)$ is a sum over
slices of the transverse gradient of the potential at comoving
distances $\chi_i$ (with sources fixed at $\chi_s$):
\begin{equation}
  \label{eq:kappa_phi}
  \kappa(\vectheta) = \frac{1}{c^2} \sum_i L_{\rm box} W(\chi_i, \chi_s) 
  \nabla^2_\vectheta \phi(\chi_i\vectheta, \chi_i).
\end{equation}
In standard general relativity, we can use the Poisson equation, the
real-space version of Eq.~(\ref{eq:phik_cont}), to obtain
\begin{equation}
  \label{eq:kappa}
  \kappa_{\rm std}(\vectheta) = \frac{3 H_0^2 \Omega_{\rm m0}}{2 c^2} \sum_i L_{\rm
    box} W(\chi_i, \chi_s) \frac{\delta^{\rm 2D}_i(\chi_i \theta)}{a(\chi_i)},
\end{equation}
where $L_{\rm box}$ is the tile boxsize, and
\begin{equation}
  \label{eq:lensing_W}
  W(\chi_i, \chi_s) \equiv \frac{\chi_i(\chi_s-\chi_i)}{\chi_s}
\end{equation}
is the weak lensing weight function.  
If the photons feel the same modified potential as the dark matter,
then we must use a modified Poisson equation such as
Eq.~(\ref{eq:phialt}) instead; note that this is not true in some
theories, e.g.\ in the fifth force-type modification of 
\citet{Nusser:2004qu}.
The converse has also been considered: \citet{White:2001kt}
constrained the parameters of a model where the potential
affecting photons was modified, but structure was assumed to have
formed as in GR.
Since we are interested in scales ($\ell \ge 100$), we use the flat-sky
approximation and define the 2D Fourier transform as
\begin{eqnarray}
  \label{eq:FT2D}
  \tilde{\kappa}(\vecl) &=& \int d\,^2\vectheta\, \kappa(\vectheta)
  e^{i\vecl \cdot \vectheta},\\
  \kappa(\vectheta) &=& \int \frac{d\,^2\vecl}{(2\pi)^2}\,
  \tilde{\kappa}(\vecl)
  e^{-i\vecl \cdot \vectheta}.
\end{eqnarray}

Assuming the mass distribution in
different redshift slices is uncorrelated, the convergence power spectrum
in this approximation is a sum over the 2D power spectra of the
projected densities:
\begin{equation}
  \label{eq:Pkappa_sum}
  \langle |\tilde{\kappa}(\ell)|^2\rangle \propto \sum_i
\frac{W^2(\chi_i, \chi_s)}{a^2(\chi_i)} 
f^2\left(k=\frac{\ell}{\chi_i}, a(\chi_i)\right)
\langle |\tilde{\delta}_i^{\rm 2D} (\ell)|^2\rangle.
\end{equation}
Here we have included the function $f(\veck,a)$, introduced in
Eq.~(\ref{eq:Gk}), that describes the modification of gravity in the
Poisson equation; for the model we consider in this paper (i.e.\ 
Eq.~(\ref{eq:phikalt})),
\begin{equation}
  \label{eq:fk}
  f(\veck,a) = 1 + \alpha\frac{1}{1
    + (|\veck| r_{\rm s}/a)^2}.
\end{equation}
We can already see that if an AG modification changes the potential
felt by dark matter and photons in the same way, then WL results can
provide a stronger constraint on AG.  In
Eq.~(\ref{eq:Pkappa_sum}), the modification will affect the two-point
function $\langle |\tilde{\kappa}(\ell)|^2\rangle$ twice: once via the
modified growth of structure from the $\langle |\tilde{\delta}_i^{\rm
  2D} (\ell)|^2\rangle$ term, and once via the $f^2(k,a)$ term that
comes directly from the modification of the potential.  We show in
Sec.~\ref{sec:sims} that for our model, these effects of AG
combine to produce a dramatic effect.

Computing the convergence power spectrum from
Eq.~(\ref{eq:Pkappa_sum}) uses only the 2D modes of the density field
at each redshift slice; a large part of the information in each
simulation box is lost by the projection.  In order to reduce
the scatter in the simulations, we included the full 3D modes in
our computation of $P_\kappa(\ell)$ from our data.  We accomplished this
by using the Limber approximation to compute $P_\kappa$ directly as
an integral over the 3D matter power spectrum $P_\delta$:
\begin{equation}
\label{eq:limber}
  P_\kappa(\ell) = \frac{9H_0^4 \Omega_{\rm m0}^2}{4c^4}
  \Delta\chi \sum_i \frac{W^2(\chi_i)}{\chi_i^2 a^2(\chi_i)}
  f^2\left(k=\frac{\ell}{\chi_i},a(\chi_i)\right)
  P_\delta\left(k=\frac{\ell}{\chi_i}, \chi_i\right).
\end{equation}
Traditionally, Eq.~(\ref{eq:limber}) is used to compute $P_\kappa(\ell)$
when one has an estimate of $P_\delta(k)$ from the halo model or
other fitting formula.  Instead we use Eq.~(\ref{eq:limber})
with $P_\delta$ measured from the PM simulations at redshifts $z_i$.
This results in a significant gain in our signal-to-noise for $P_\kappa$
as shown in Fig.~\ref{fig:2dmaps}: the error bars in the
standard method based on measuring $P_\kappa$ from $\kappa$ maps are
significantly larger than in our method.

\begin{figure*}[t]
  \centering
  \includegraphics[width=0.425\linewidth,height=0.425\linewidth]{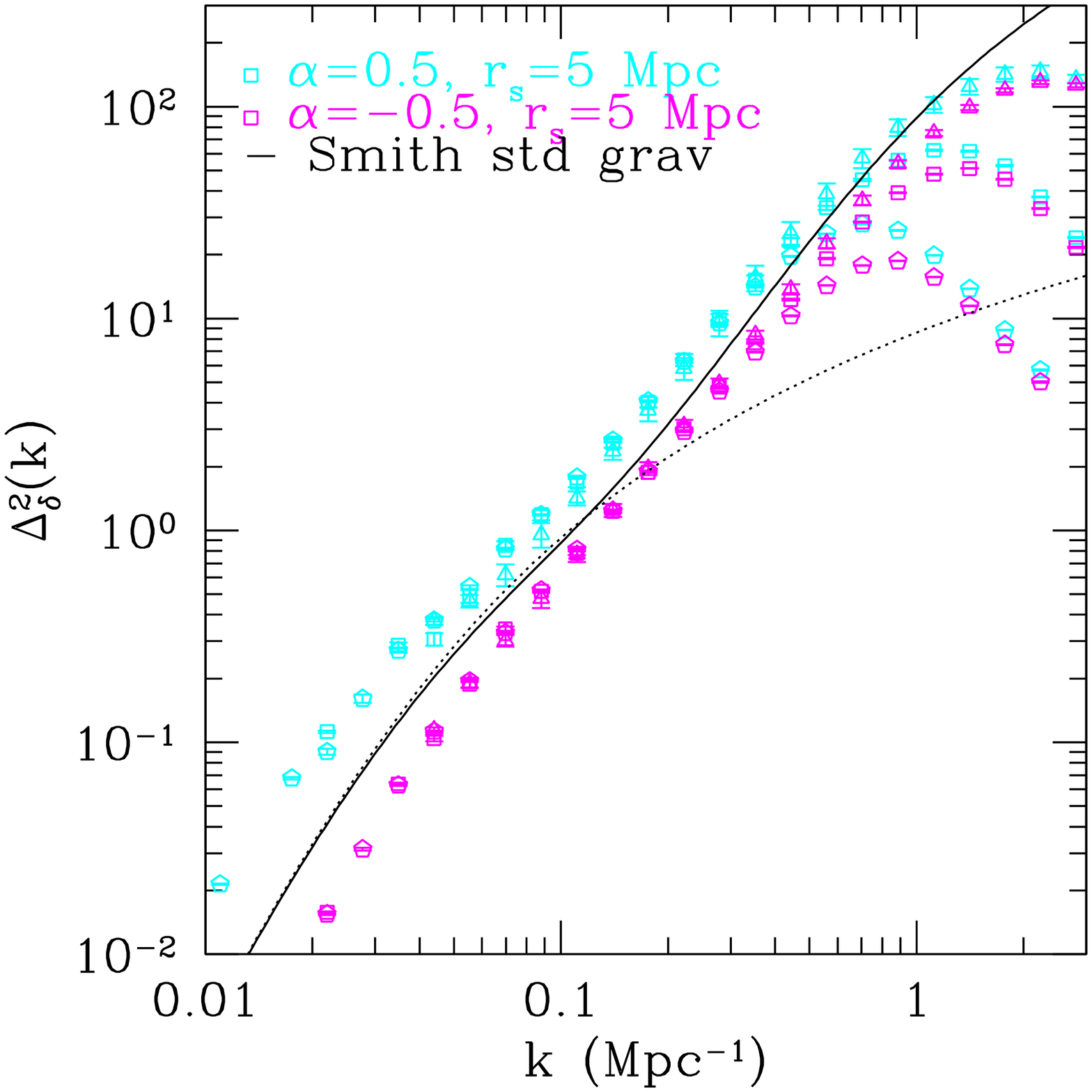}
  \includegraphics[width=0.425\linewidth,height=0.425\linewidth]{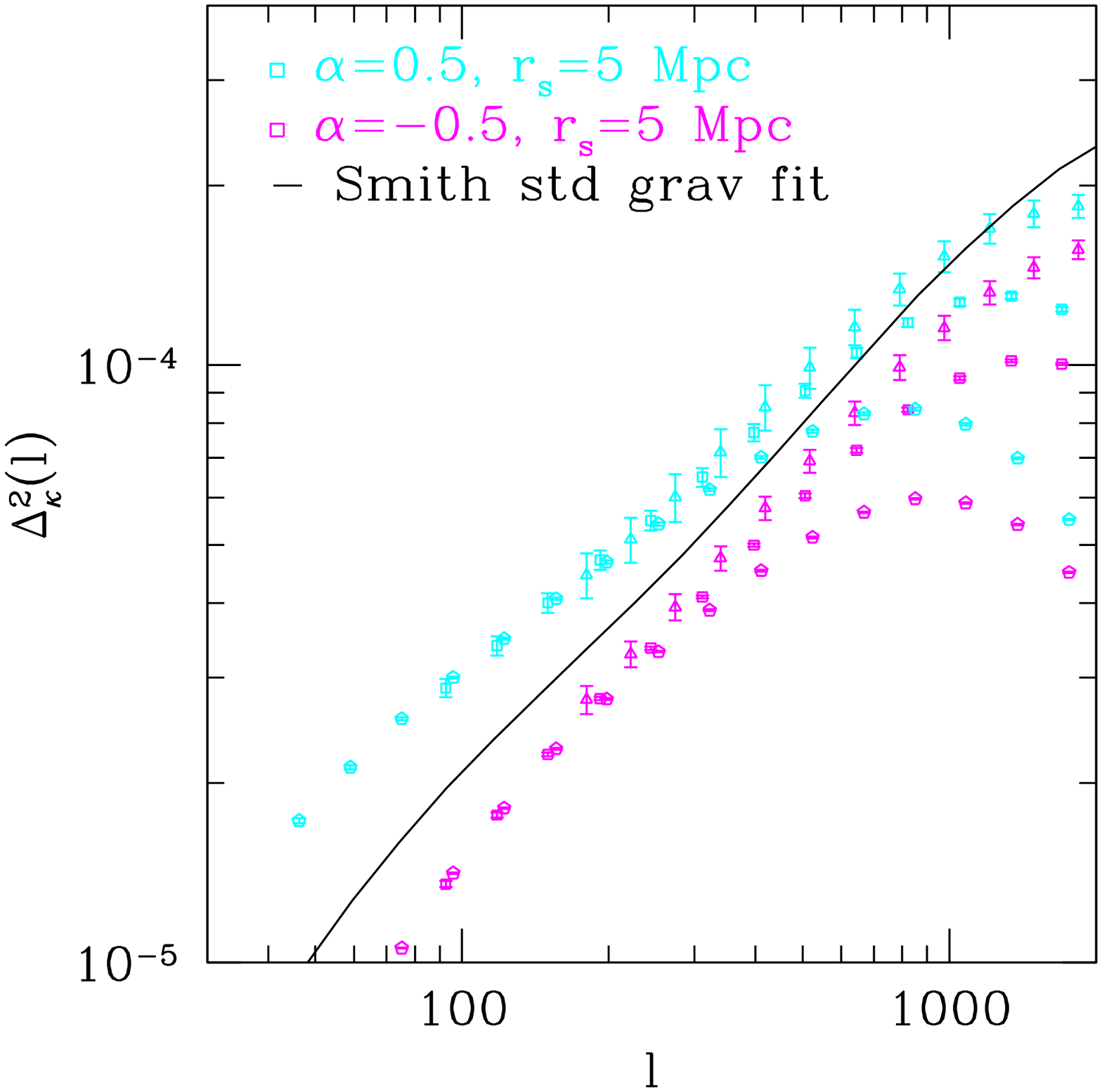}  
  \caption{\small Dimensionless 3D power and convergence power spectra
    from simulation.  The colored pentagons are simulation data
    produced using a 400h$^{-1}$~Mpc box, the squares have a
    200h$^{-1}$~Mpc box, and the triangles have 100h$^{-1}$~Mpc (which
    we use in the main portion of our simulations). The simulations
    are consistent with each other up to their resolution limit,
    regardless of whether or not $r_s$ is initially outside the box.
}
  \label{fig:boxscale_check}
\end{figure*}

For each realization we obtain the power spectrum in $100h^{-1}$~Mpc
boxes tiled along the line of sight.  Using the discrete version of
Equation~(\ref{eq:limber}) we get $P_\kappa(\ell)$ for each
realization from $P_\delta(k)$ binned in spherical shells in
wavenumber $k$; the bins we need at each redshift are given by
$k=\ell/\chi_i$.  Eight realizations give us our estimate of the
scatter in our results.  We take our results to be valid up to
wavenumber $k\approx 1$~Mpc$^{-1}$, which is about $k_{\rm nyq}/2$;
for $P_\kappa$ this corresponds to $\ell\approx 1000$.  Additionally,
we have checked the validity of our modified gravity simulations for
larger boxsizes.  Since at the start of our simulation, the comoving
scale $r_s(z=50)\approx 250$~Mpc is larger than $L_{\rm box} =
100$h$^{-1}$~Mpc, but at $z=0$, $r_s \ll L_{\rm box}$, we needed to
insure that no numerical artifacts are produced when $r_s$ crosses the
box scale.  In Fig.~\ref{fig:boxscale_check}, we find that this is
indeed the case.

\begin{figure*}[t]
  \centering
  \includegraphics[width=0.35\linewidth,height=0.425\linewidth]{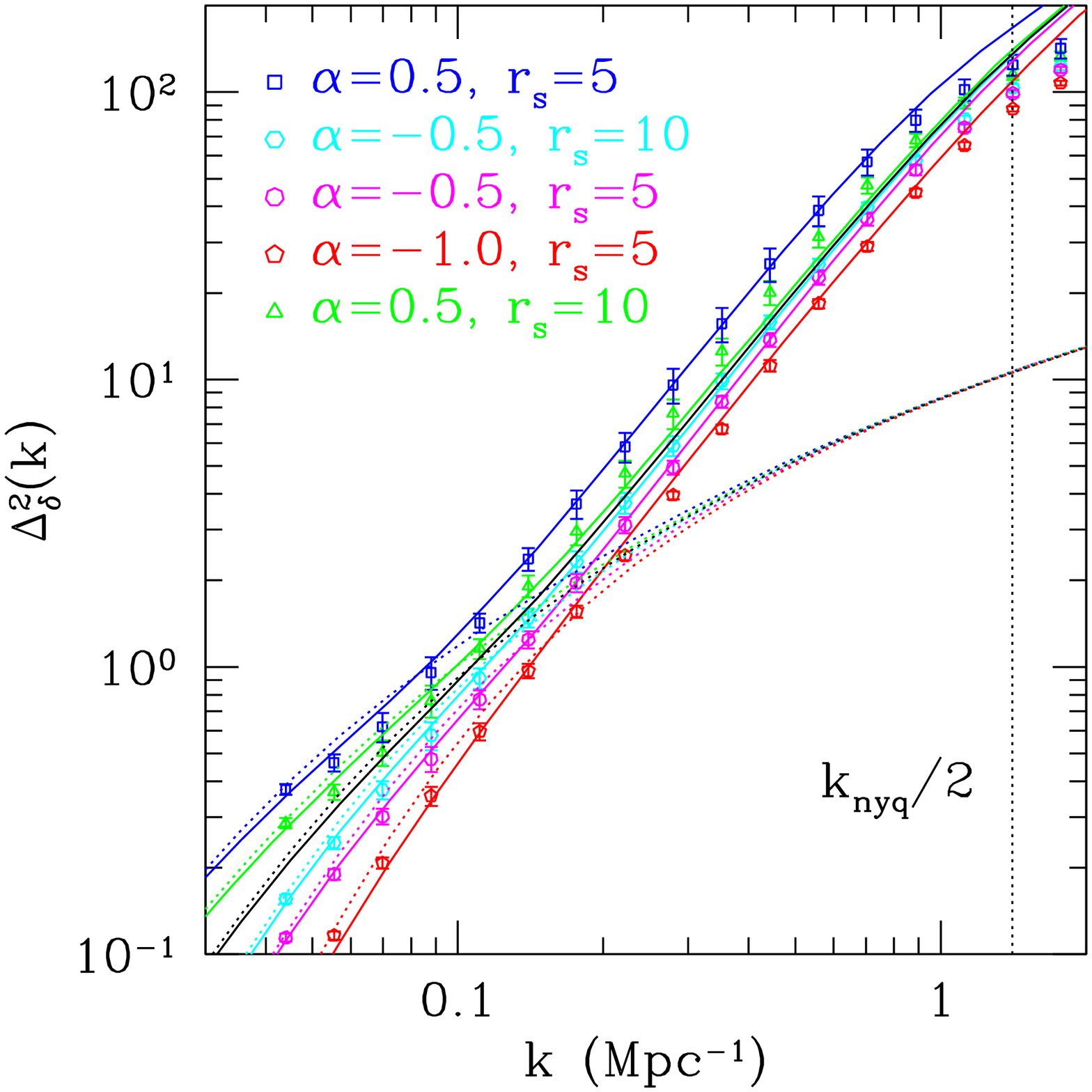}
  \hfil
  \includegraphics[width=0.5\linewidth,height=0.425\linewidth]{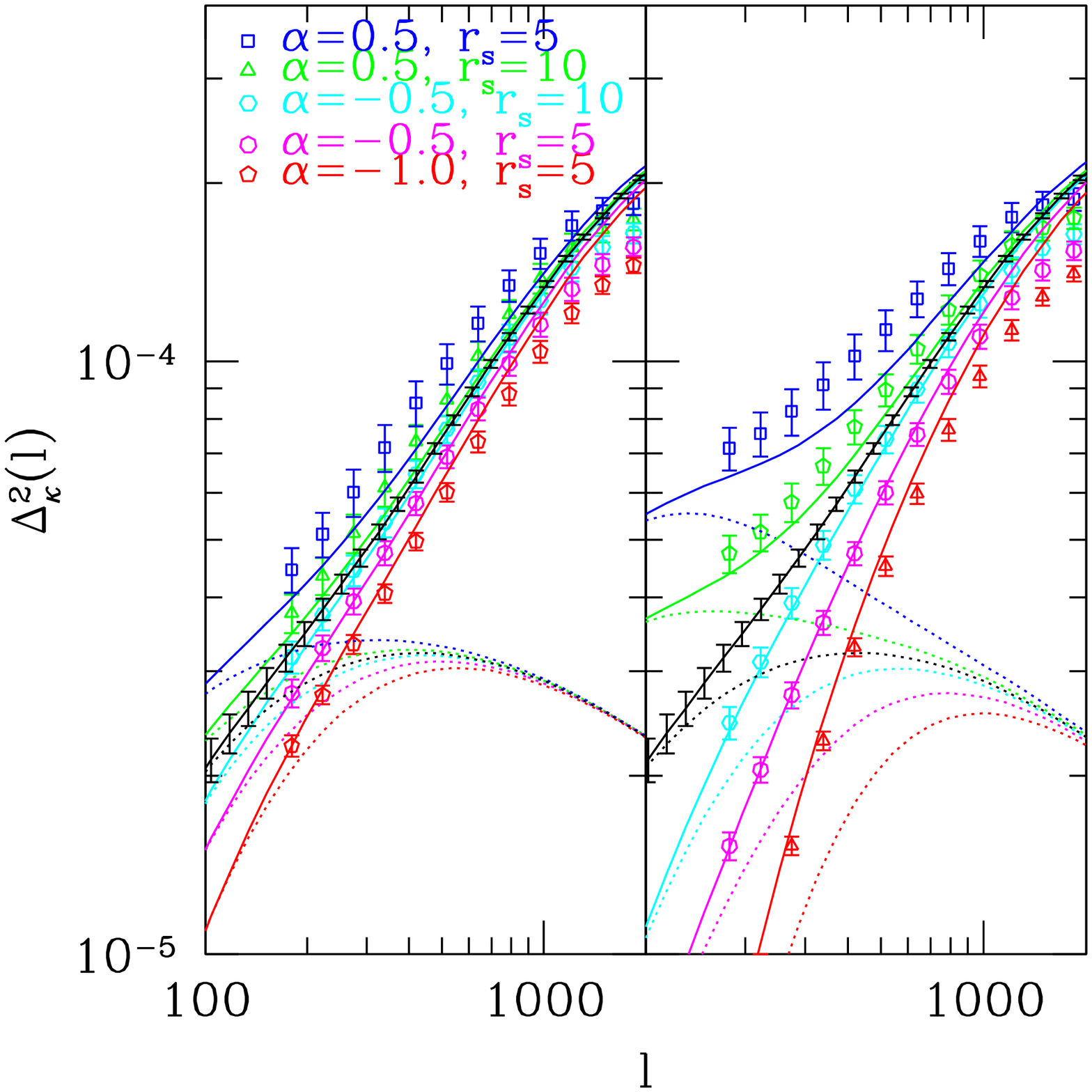}
  \caption{\small Dimensionless 3D power ($\Delta^2_\delta(k) =
    k^3P(k)/2\pi^2$) and convergence power ($\Delta^2_\kappa(\ell) =
    \ell^2P_\kappa(\ell)/2\pi$) for standard and alternative gravity
    models.  The black curves for standard gravity are computed using
    the \citet{Peacock:1996ci} fitting formula for the nonlinear 3D
    power, and using it in the Limber integral for $P_\kappa$.  
    The dotted curves are predictions from linear
    theory.   The solid curves are our
    analytic fits for the nonlinear spectra of modified gravity models (as
    indicated by the legends), while the symbols show simulation
    measurements.  
    The
    two subpanels for $\Delta^2_\kappa$ illustrate two choices for how WL power
    spectra are affected by a modified potential: the
    curves in the left subpanel are computed assuming a GR deflection
    law, i.e.\ setting $f^2 = 1$ in Eq.~(\ref{eq:limber}),
    whereas those on the right are computed with a modified potential
    for photons.  The error bars on the solid curves
    in the right panels show the expected statistical errors from a
    future lensing survey (see text in Section 4 for details). }
  \label{fig:overplot}
\end{figure*}

\begin{figure*}[t]
  \centering
  \includegraphics[width=0.35\linewidth,height=0.425\linewidth]{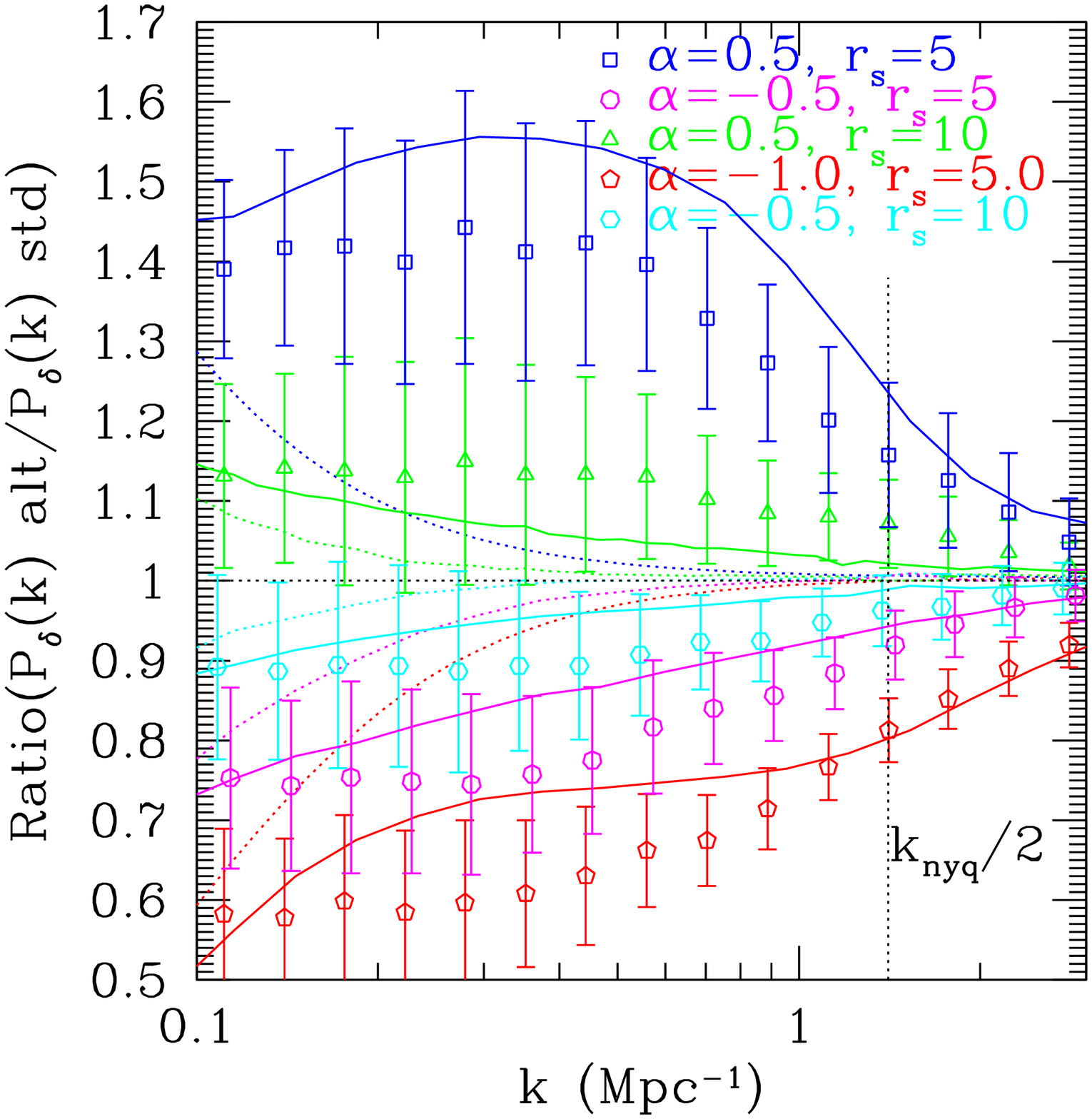}
  \hfil
  \includegraphics[width=0.5\linewidth,height=0.425\linewidth]{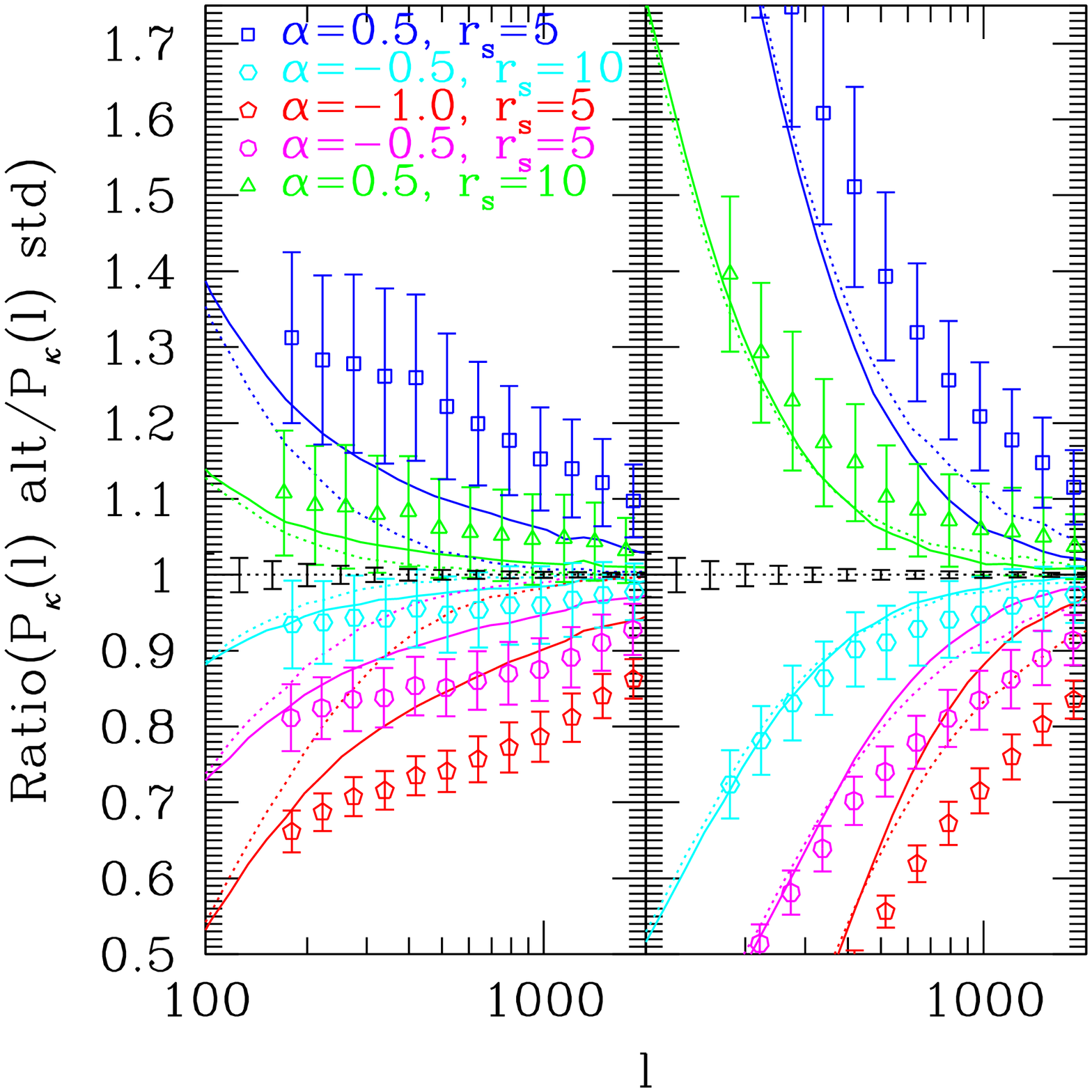}
  \caption{\small Ratio plots of dimensionless 3D power ($\Delta^2_\delta(k)=
    k^3P(k)/2\pi^2$) and convergence power ($\Delta^2_\kappa(\ell) =
    \ell^2P_\kappa(\ell)/2\pi$). The solid curves show the ratio 
    of the Peacock-Dodds prediction for AG with standard
    gravity. The dotted curves are the linear theory
    ratios.  We divided our AG simulation power spectra by the
    corresponding points from the standard gravity simulation to obtain
    the points. 
    As in Fig.~\ref{fig:overplot}, the left $P_\kappa$ subpanel shows
    the WL result for the GR deflection law, while the right subpanel
    shows that the deviation of AG from the standard gravity result is
    substantially enhanced if the deflection law is modified along
    with the growth of structure.
    The error bars on the horizontal line in the right panels show the
  expected statistical errors from future lensing surveys (see text
  for details). 
  }
  \label{fig:ratio}
\end{figure*}

\section{Simulation Results} 
\label{sec:sims}

We have run ensembles of simulations using standard gravity and the AG
potential given by Eqs.~(\ref{eq:phialt}, \ref{eq:phikalt}) with five
different sets of parameters: $\alpha = 0.5, r_{\rm s} = 5$~Mpc;
$\alpha = -0.5, r_{\rm s} = 5$~Mpc; $\alpha = 0.5, r_{\rm s} =
10$~Mpc; and $\alpha = -0.5, r_{\rm s} = 10$~Mpc.  These sets of
parameters are within the $2\sigma$ range of constraints set by
\citeauthor{Shirata:2005yr} using SDSS data; the last two models given have
the smallest deviation from standard gravity, for these models the
linear spectrum differs by 20\% at $k=0.05$~Mpc$^{-1}$.  We also
consider a model that has significantly less power on large scales:
$\alpha=-1.0, r_s = 5$ Mpc.

The left panel of Fig.~\ref{fig:overplot} shows the 3D power for
standard gravity and the five AG models, while the left panel of
Fig.~\ref{fig:ratio} shows them as ratios to standard gravity.  There
is a statistically significant difference between the models at the
smallest scales resolved by our simulations, where the general trend
is that models with larger $|\alpha|$ and smaller $r_s$ are more
different from standard gravity.  The overall shape of the linear and
nonlinear AG spectra remains similar to the shape of the standard
gravity spectra; as expected, the positive $\alpha$ models have excess
power on large scales, while the models with negative $\alpha$ have
less large-scale power compared to standard gravity.  The nonlinear
scale of each of the models is around $k=0.1$--0.2~Mpc$^{-1}$, similar
to standard gravity; on scales smaller than this, the linear spectra
asymptote to the standard gravity linear curve.  However, the 3D
nonlinear spectra show clearly that changing gravity on large scales
propagates into the nonlinear regime: in the nonlinear region
$k=0.5$--1.0~Mpc$^{-1}$, where the linear spectra are within a few
percent of standard gravity, the nonlinear 3D spectra differ by 10\%
or more.

The $P_\kappa$ plots in
Figs.~\ref{fig:overplot},\ref{fig:ratio} are split up into two
subpanels: in the left subpanel, photons are not affected by the modification
of the potential (but the growth of structure is still altered),
corresponding to setting $f^2 = 1$ in Eq.~(\ref{eq:limber}),
while in the right panel photons feel the modified potential.  Each
$P_\kappa$ is given by the limber integral over the corresponding
$P_\delta$ in the left panel, the difference is only whether the $f^2$
term in Eq.~(\ref{eq:limber}) is included. If light deflection and
structure formation are both affected by the modified potential, then
the difference between the $P_\kappa$ subpanels in
Figs.~\ref{fig:overplot},\ref{fig:ratio} shows that an AG model would
be much more easily ruled out.  Because the modified potential for
photons and the modified growth of structure can reinforce each other,
WL statistics potentially have more power to constrain AG theories
than large-scale structure observations (such as the galaxy power
spectrum) do: in the left panel of Fig.~\ref{fig:ratio}, we see that
the cyan and green models ($|\alpha|=0.5$, $r_s=10$~Mpc) are perhaps
just barely detectable using the power spectrum; but if light
deflection as well as large-scale structure is modified, these models
are easily ruled out.

For $P_\kappa$ as well the differences from standard gravity extend to
smaller angular scales; at our resolution limit of $\ell \sim 1000$,
there are still observable differences at the level of several
standard deviations for our fiducial survey.  The
error bars on the black lines in each of the right panels are the
statistical errors from a hypothetical lensing survey with $f_{\rm
  sky}=0.1$ and $n_{\rm gal}=40$~arcmin$^{-2}$.  The general trends are
the same as for $P_\delta$: the models with larger $|\alpha|$ and
smaller $r_s$ exhibit larger deviations from standard gravity, and the
linear and nonlinear spectra are more different on large angular
scales.  As is evident from Fig.~\ref{fig:overplot}, while the linear
$P_\kappa$ asymptotes to the the standard gravity result on small
scales, in the observable regime ($\ell > 100$), $P_\kappa$ has
nonlinear contributions.  
Here we have not performed a
full parameter analysis that would involve CMB priors and variations
of all relevant parameters.

\begin{figure*}[t]
  \centering
  \includegraphics[width=0.425\linewidth]{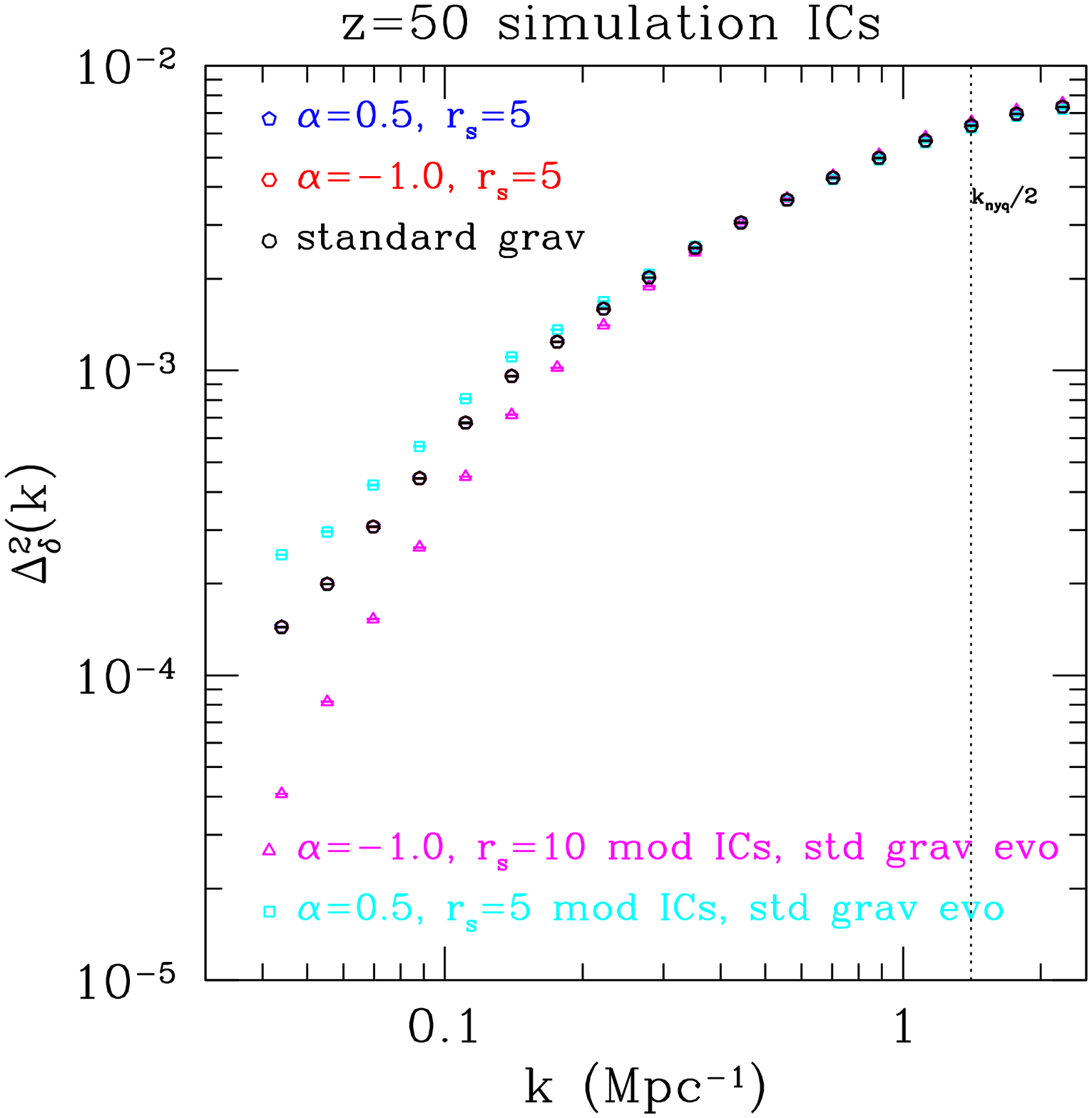}
  \hfil
  \includegraphics[width=0.425\linewidth]{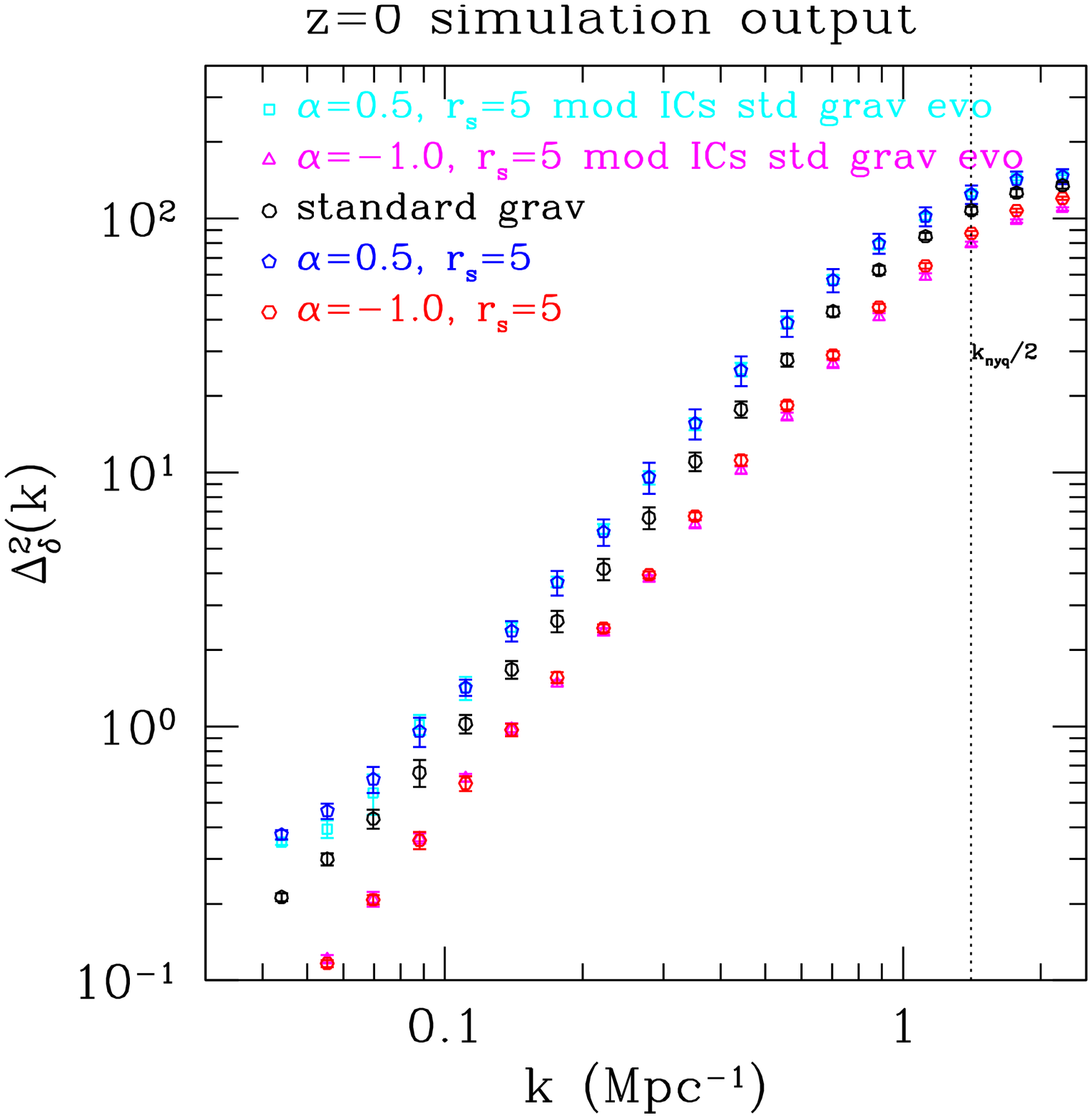}
  \caption{\small Comparing the effect of changing the initial
    conditions to changing the potential.  The ICs at $z=50$ are on
    the left, and the $z=0$ outputs are on the right.  The cyan and
    magenta points are simulations that have a modified initial
    spectrum shape but are evolved with the standard gravity
    potential, while the red and blue points are evolved using the AG
    potential with the correct initial spectrum (as in the rest of
    the paper).  
    The two pairs of points for each model (blue and cyan points, and
    the red and magenta points), are within errors of each other at the
    end of the simulation (the $z=0$ right panel). 
    Hence there is an approximate degeneracy between the
    shape of the initial power spectrum and the shape of the
    potential during its evolution under gravity.}
  \label{fig:altICs}
\end{figure*}

\subsection{Analytical Approximations}
\label{sec:PD}

From the power spectra in Fig.~\ref{fig:overplot} and
Fig.~\ref{fig:ratio}, one cannot tell whether the observed
differences on small scales at late times are a result of the changed
non-linear evolution in the AG models, or whether the structure formed
under the influence of normal gravity and merely started from
different initial conditions.

To answer this question we ran simulations whose
initial conditions had the same shape as the late-time AG linear power
spectrum, but were evolved using the standard gravity potential.  The
results, shown in Fig.~\ref{fig:altICs}, are striking: the 3D power
spectra of simulations, which had the $z=0$ linear shape of the AG
model at the initial time, came out the same as the regular AG runs.  
Recall that since $r_s$ is a fixed physical length scale in the
AG model, at the start of the simulation at $z=50$, it is $250-500$Mpc
in comoving coordinates. This is larger than the simulation box size of
$100h^{-1}$~Mpc. So we would expect the AG power spectrum at $z=50$
to be identical to that of standard gravity.

The plots in Fig.~\ref{fig:altICs} show a kind of universality in cold
dark matter structure formation: the way non-linear structures form is
not uniquely determined by specifying the detailed shape of the
potential.  Our simulations show that the DM power spectrum by itself
cannot distinguish between changing the shape of the initial power
spectrum and changing the shape of the gravitational potential.

\begin{figure*}[t]
  \centering
  \includegraphics[width=0.425\linewidth]{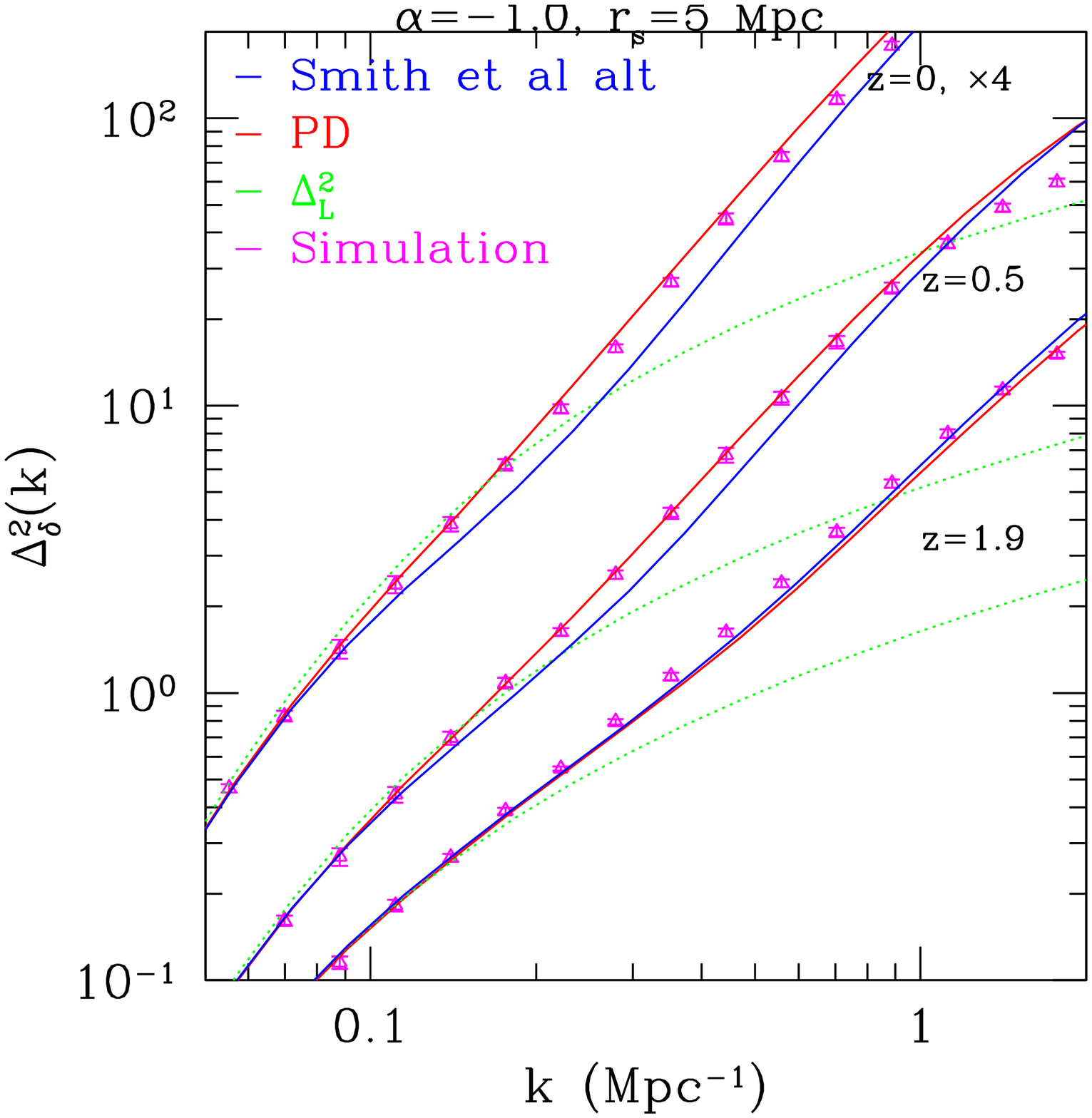}
  \hfil
  \includegraphics[width=0.425\linewidth]{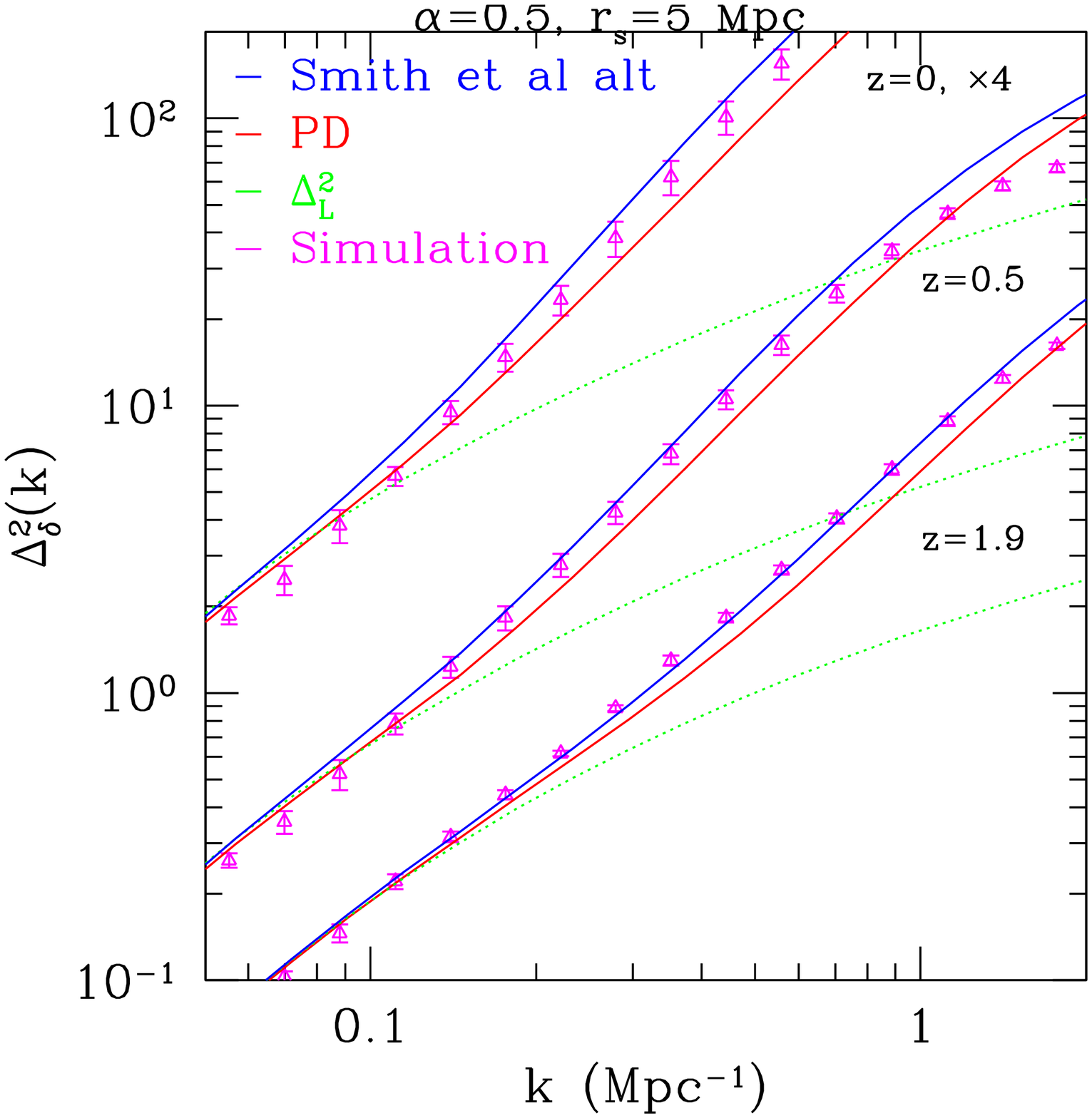}
  \caption{\small Dimensionless power for negative and positive
    $\alpha$ models at different redshift, compared to the
    \citet{Smith:2002dz} (blue, dashed) 
    and \citet{Peacock:1996ci} (red, solid) fitting formulae.
    The $z=0$ outputs have been translated by a factor of 4 in the
    $y$-direction for legibility.  For negative $\alpha$, PD fits
    better than Smith, while for positive $\alpha$, the simulation
    points lie between the two predictions.
  }
  \label{fig:smithPD}
\end{figure*}

This is further revealed by testing the power spectra measured in
simulations against analytical fitting functions that have been
calibrated for standard gravity.  The Peacock-Dodds (PD) formula
\citep{Peacock:1996ci}, using a 
mapping of length scales between the linear and nonlinear regime,
gives the dimensionless nonlinear 3D power spectrum
$\Delta^2_{\delta}$ as a function of the dimensionless linear power
$\Delta^2_{\rm \delta,L}$ to around 10\% accuracy (when compared to
simulations).  \citet{Shirata:2005yr} use the PD formula to extend
their results to non-linear scales.

We have tested the PD formula with AG simulations
by replacing the standard gravity linear power spectrum in the formula
with the linear spectrum from the AG model.  The results for $P_\delta$
are shown in the left panel of Fig.~\ref{fig:overplot}, in which the dashed
lines are produced using the Peacock-Dodds formula.  The right panel
shows $P_\kappa$ fits generated by integrating the 3D power along the
line of sight using the Limber approximation.  The fits in
Fig.~\ref{fig:overplot} are accurate at about the 10\% level, with the
accuracy apparently better for models closer to standard gravity. 
From the ratio plots in Fig.~\ref{fig:ratio}, we also see that the
positive $\alpha$ model (more large scale power) does worse. 

The fitting formula of \citeauthor{Smith:2002dz}\ performs the same task as
the PD formula in that, given a wavenumber $k$ and redshift $z$, it
provides an estimate of the non-linear power spectrum
$\Delta^2_\delta(k,z)$.  
The Smith formula is inspired by the halo
model: it breaks up the non-linear power into two pieces, a
quasi-linear term and a one-halo like term. 
We adapted the Smith formula for use with alternative
gravity by using the AG linear spectrum, 
but we see in Fig.~\ref{fig:smithPD} that PD is a much better
fit to the data for negative $\alpha$ (for the
positive-$\alpha$ model that we have tested, the Smith and PD fits
are comparable at low redshift).  This may be due to the use of two
separate terms in the Smith formula, one of which is calibrated on the
basis of the shapes of CDM halos in standard gravity. 
It was designed and tested for  simulations with scale-free
initial spectra or CDM initial conditions, but not for initial spectra
with a very different shape, such as those produced by a
scale-dependent growth factor, initial conditions with a shape like
those in Fig.~\ref{fig:altICs}.

\begin{figure*}[t]
  \centering
  \includegraphics[width=0.425\linewidth]{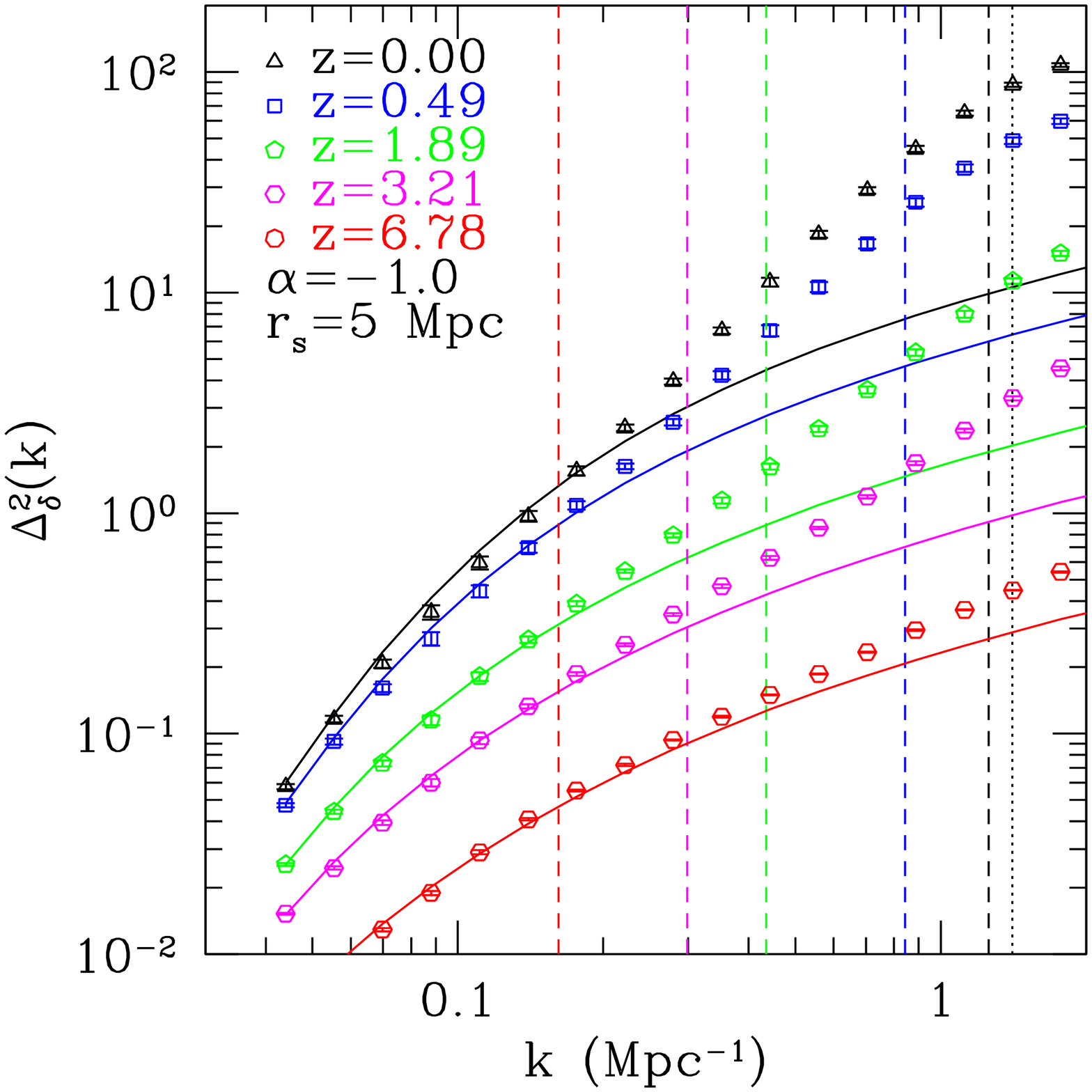}
  \hfil
  \includegraphics[width=0.425\linewidth]{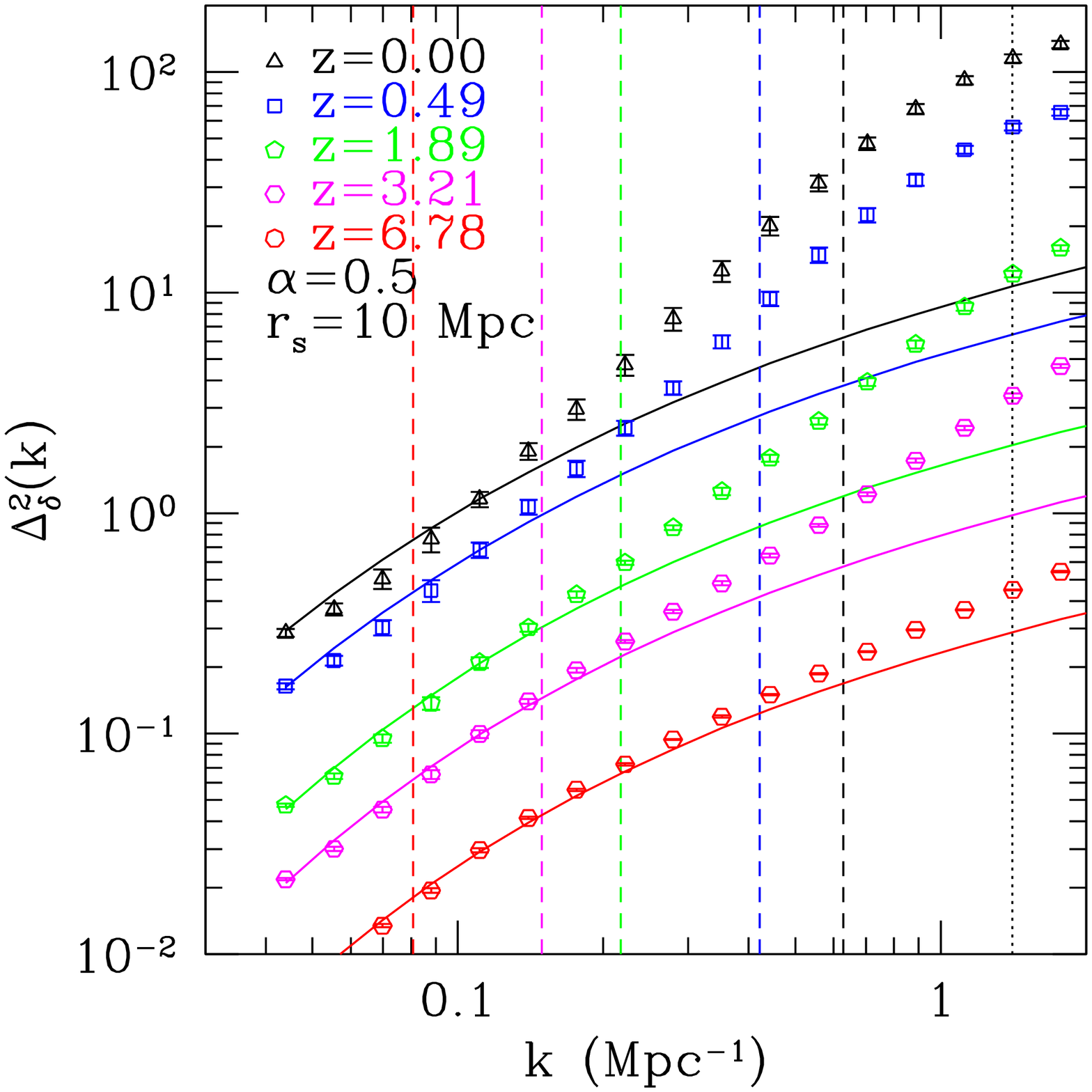}
  \caption{\small Dimensionless 3D power spectrum ($\Delta^2_\delta(k,z)=k^3
    P(k,z)/2\pi^2$) as a function of wavenumber and redshift for two
    different modified gravity models.  The solid lines are the linear
    spectra while the points and error bars are the average of eight
    realizations from simulation; the vertical long-dashed lines
    indicate the value of $k_s(z) = 2\pi a(z)/r_s$, which is the
    comoving wavenumber corresponding to the physical scale $r_s$.
    The nonlinear scale $k_{\rm nl}(z)$ can be read off the graph by
    observing where the linear and nonlinear power spectra begin to
    diverge.  We can see that in each model, $k_s < k_{\rm nl}$ initially,
    but at $z=0$, $k_s > k_{\rm nl}$ in each case.}
  \label{fig:alpha_models_rs_z}
\end{figure*}

To summarize, our results suggest
that the nonlinear power spectrum in alternative gravity models 
is captured completely by
the change in the linear growth factor. This result is consistent with
the approach used in the PD fitting function. A similar result is shown in
the recent study of \citet{Linder:2005hc},
who found that nonlinear spectra for a class of dark energy models can
be accurately described by appropriate choice of length and time
scales. We must emphasize, however, that our results are only valid
for a certain range of scales.  We can estimate over what range our
results should be valid by examining what happens to the comoving
scale $r_s$ and the nonlinear scale as the simulation progresses.  The
comoving wavenumber corresponding to $r_s$ is defined by $k_s(z) =
2\pi / ((1+z) r_s)$.  For a $r_s=5$~Mpc model,
Fig.~\ref{fig:alpha_models_rs_z} shows how the wavenumber $k_s$ ranges
from approximately 1.3~Mpc$^{-1}$ at the present to 0.16~Mpc$^{-1}$ at
$z=6.78$.  Inspection also reveals that the nonlinear wavenumber
$k_{\rm nl}$ (the scale where the linear and nonlinear spectra begin
to diverge) ranges from approximately $k_{\rm nl}=0.2$~Mpc$^{-1}$ at
the present to $k_{\rm nl}=0.5$~Mpc$^{-1}$ at $z=6.78$ for the
$\alpha=-1$, $r_s=5$~Mpc model.  So we see that $r_s$ actually starts
out larger than $r_{\rm nl}$, but ends up well inside the nonlinear
scale as time goes by; the redshift where they coincide depends on the
values of $r_s$ and alpha chosen, for $\alpha=-1$, $r_s=5$~Mpc one can
see that it occurs at about $z \approx 3.2$, whereas for the
$\alpha=0.5$, $r_s=10$~Mpc model, the scales cross later at a redshift
of around $z \approx 1.9$.  As long as the boxsize and resolution of
our simulations can adequately capture the dynamics of $r_s$ and
$r_{\rm nl}$ over a range of redshifts, i.e.\ for any model where the
nonlinear scale and $r_s$ cross well before redshift $z=0$, the
conclusions we reach from our simulation should be valid; we estimate
that this would be true for a range of $r_s$ at least from about 1~Mpc
to 20~Mpc.

It is interesting that quasilinear perturbation theory 
\citep{Jain:1993jh,Bernardeau:2001qr} 
would have suggested some departures from
this universality: equations for the 
second order density and velocity fields contain the linear fields as
well as the $\nabla \phi$ term in the Euler equation. So the dependence 
of the second order
terms on the scale dependent function $f(k,t)$ introduced in
Eq.~(\ref{eq:Gk}) would not be completely determined by the linear
solution. Our results show that this quasilinear departure 
is likely very small. 
Generally speaking, the halo model description, assuming it
describes alternative gravity models, is consistent with universality:
halo bias, the linear spectrum and halo mass function 
depend only on the linear growth and
initial power spectrum. Since the small scales where halo
structure may play a role are not well probed by our simulations, it
is not surprising that we find an approximate universality in the
nonlinear power spectrum. The PD formula gives results close to
the halo model for $\Lambda$CDM \citep{Smith:2002dz}.

\section{Discussion}
\label{sec:discussion}

We have performed N-body simulations of large scale
structure formation with a modified Newtonian potential in a
$\Lambda$CDM background. This is intended to approximate 
alternative gravity theories that are designed to match the observed
acceleration of the universe. We focus on the quasilinear and 
non-linear regime of clustering at low redshift. 
Our simulations resolve the 3D power spectrum of matter on  
scales of $k\approx 0.05$--1.0~Mpc$^{-1}$.  We used the 3D simulations
to compute the weak lensing power spectrum over angular wavenumbers 
$\ell\approx 100$--1000. We used a technique for this that reduces the
scatter in measurements from simulations (described in Section
3.3). The range of scales we studied is expected to be
observable with high accuracy with planned surveys. 
The nonlinear modification of the power
spectra ranges from $10\%$ effects in the quasilinear regime to an
order of magnitude at the small scale end. While the accuracy of our
simulated spectra is typically $10\%$ over this range, the relative
accuracy for predictions of different gravity models is significantly
better. 

We find that nonlinear effects propagate the
difference between the power spectra to scales where the linear spectra are
nearly identical. This is the expected effect of mode
coupling in nonlinear gravitational evolution \citep{Jain:1993jh}.
The result is that at scales of $k\approx 0.5$~Mpc$^{-1}$,
the modified gravity power spectra differ by over 10$\%$, while the
linear spectra are within 5$\%$. Similar differences are seen in the
weak lensing spectra at $\ell\approx 500$, and if the potential for
photons is modified in a similar way as the potential for matter, the
effects reinforce each other and the differences in the weak lensing
spectra are much larger. These scales are of great
interest because the observational errors are expected to be small and
theoretical interpretation can be made without modeling of
non-gravitational effects (at least for the lensing spectra). 
We compare the differences between models 
to the expected statistical errors from a wide area lensing 
survey to show that it should be possible to directly constrain 
the parameters of an alternative gravity scenario. Models within the 
2-$\sigma$ limits of current galaxy surveys would be easily distinguished 
by future lensing measurements. A detailed study of this is left for 
future work. 

Our results are consistent with a universality in nonlinear gravity,
which makes the nonlinear power spectrum a function only of the
initial (Gaussian) conditions and linear growth, for modifications on
length scales for which there are likely to be precise observations in
the near future.  We find that the lensing and 3D power spectra cannot
distinguish between simulations which were started with appropriately
modified initial conditions and evolved with standard gravity, and
those which were evolved with a modified gravitational potential.
This means that for observationally accessible scales, there is a
degeneracy between the shape of the potential used to evolve the
simulation and the shape of the initial conditions. For simple
modified gravity models, it means that another constraint on the
primordial power spectrum (such as the CMB), or measurements at
multiple redshifts, must be used to test for modified gravity. It may
also be that the structure of dark matter halos, not well probed by
our simulations, are different for AG models. This is an interesting
topic for future work with higher resolution simulations.  The
skewness or bispectrum may also help distinguish alternative gravity
models, as suggested by \citet{Bernardeau:2004ar} and
\citet{Sealfon:2004gz}, since its dependence on the scale dependent
function $f(k,t)$ in Eq.~(\ref{eq:Gk}) is different.

We tested analytical approximations to the nonlinear spectra for
modified gravity models. We found that while the
Peacock-Dodds fitting formula was accurate to 10--20\% in comparison
to the simulated spectra, its relative accuracy between different
models is significantly better and within the errors of our 
measurements. The Smith et al.\ formula does somewhat worse for the
models studied. 

Our simulations are a useful first step in studying the effects that
an alternative gravity model has on large scale structure formation.
It would be of great interest to simulate the 
dynamics of a full alternative gravity theory, 
instead of our approach (which may be a convenient approximation 
but is not even covariant).  However, a full alternative gravity theory 
that one can simulate is hard to come by; e.g. going beyond the 
linear regime for the DGP model \citep{Lue:2005ya}, even in the 
quasilinear regime, is an unsolved problem.


\acknowledgments Our N-body simulations used the code kindly made
public by Anatoly Klypin. We thank Derek Dolney for his help and contributions
in running the numerical simulations. We acknowledge helpful
discussions with Eric Linder, Mike Hudson, Andrey Kravtsov, Carolyn Sealfon, 
Roman Scoccimarro, Ravi Sheth, 
Robert Smith and Masahiro Takada. This work is supported in part by NASA
grant NAG5-10924 and and NSF grant AST03-07297. 

\bibliographystyle{apsrev}

\end{document}